\documentclass{iopart}

\usepackage{graphicx} 
\usepackage{verbatim} 
\usepackage{bm}
\usepackage{color}
\usepackage[colorlinks=true]{hyperref}

\begin{document}


\def\snn{\mbox{$\sqrt{s_{_{\rm NN}}}$}}   
\def\pt{\mbox{$p_{\rm T}$}}
\newcommand{\gapp}{\,{\raisebox{-.2ex}{$\stackrel{>}{_\sim}$}}\,}
\newcommand{\lapp}{\,{\raisebox{-.2ex}{$\stackrel{<}{_\sim}$}}\,}
\newcommand{\dla}{\langle\!\langle}
\newcommand{\dra}{\rangle\!\rangle}
\newcommand{ \ds }{\displaystyle}


\title{Collective Expansion at the LHC: selected ALICE anisotropic flow measurements}

\markboth{Collective Expansion at the LHC}{Snellings}

\author{Raimond Snellings$^{1,2}$}
\address{
$^1$ 
Utrecht University, 
Princetonplein 5, 
3508 TA Utrecht, 
The Netherlands. \\
$^2$ 
Nikhef, 
Sciencepark 105, 
1098 XG Amsterdam, 
The Netherlands.
}
\ead{Raimond.Snellings@cern.ch}


\begin{abstract}

The collective expansion of matter created in collisions of heavy-ions, 
ranging from collision energies of tens of MeV to a few TeV per nucleon pair, 
proved to be one of the best probes to study the detailed properties of these unknown states of matter. 
Collective expansion originates from the initial pressure gradients in the created hot and dense matter. 
These pressure gradients transform the initial spatial deformations and inhomogeneities of the created matter into 
momentum anisotropies of the final state particle production, which we call anisotropic flow. 
These momentum anisotropies are experimentally characterised by so-called flow harmonics.
In this paper I review ALICE measurements of the flow harmonics at the CERN 
Large Hadron Collider and discuss some of the open questions.    

\end{abstract}

\maketitle

\noindent

\section{INTRODUCTION}
\label{introduction}
In the world around us, quarks and gluons do not exist as free particles because they are permanently bound 
into hadrons by the strong interaction. 
At very high temperatures and densities however, hadronic matter is expected to undergo a phase transition to a 
new state of matter, the Quark-Gluon Plasma (QGP), where quark and gluon degrees of freedom are not 
anymore confined inside the hadrons~\cite{Wilson:1974sk, Shuryak:1980tp, Chapline:1974zf, Lee:1974ma, Lee:1978mf,
Collins:1974ky, Pisarski:1983ms,Rajagopal:2000wf}. 
Such temperatures and densities were present in the early Universe until the first microseconds after the Big Bang. 
Not only in the early Universe, but also in heavy-ion collisions at ultra-relativistic energies it is possible to reach 
temperatures and energy densities well above the phase-transition point. 
The highest temperatures and energy densities are currently reached in collisions of heavy-ions at the 
Large Hadron Collider (LHC). 

Measurements of the collective expansion of the QGP, in particular the azimuthal anisotropies in the 
expansion~\cite{Ollitrault:1992bk}, 
provide us with constraints on the properties of the system such as the equation of state (EoS)~\cite{Huovinen:2009yb} 
and transport coefficients like the kinematic viscosity, defined as the shear viscosity over entropy ratio ($\eta$/s). 
The collective expansion, also called flow, originates from the initial pressure gradients in the created hot and dense matter. 
These pressure gradients transform the initial spatial deformations and inhomogeneities of the created matter into 
momentum anisotropies of the final state particle production, which are experimentally characterised by so-called flow harmonics~\cite{Voloshin:1994mz}. For recent reviews see~\cite{Huovinen:2006jp,Voloshin:2008dg,Teaney:2009qa,Heinz:2013th,Gale:2013da,Luzum:2013yya}

Based on asymptotic freedom~\cite{Gross:1973id,Politzer:1973fx} in Quantum Chromo Dynamics (QCD) and color Debye 
screening~\cite{Nadkarni:1986cz}, the properties of the QGP were 
expected to be similar to a weakly interacting gas of quarks and gluons. In that case the mean free path 
in the QGP is large, which implies a large viscosity. A direct consequence of a large viscosity is that the 
system will not develop strong collective expansion. 
Therefore the discovery of a very large $2^{nd}$ order flow harmonic~\cite{Ackermann:2000tr,Adler:2003kt}, 
called elliptic flow, at the Relativistic Heavy Ion Collider (RHIC) and more recently at the 
LHC~\cite{Aamodt:2010pa,ATLAS:2011ah,Chatrchyan:2012ta}  
changed dramatically our understanding of the QGP.

Currently, the buildup of this large flow is theoretically best understood in a relativistic 
viscous hydrodynamic model description. 
In these models the large flow is understood by assuming that the system is, very quickly after the collision, 
in local equilibrium and forms a strongly coupled quark-gluon liquid. 
Only after it was realised that already a small kinematic viscosity 
significantly reduces the buildup of the elliptic flow~\cite{Teaney:2003kp,Romatschke:2007mq,Luzum:2008cw}, 
it became clear that the system 
produced behaves as an almost ideal or inviscid fluid, which is a fluid that has almost no resistance to shear stress. 

While lattice QCD calculations can calculate precisely quantities like the EoS of the QGP, 
they unfortunately currently provide little guidance to dynamical quantities like the kinematic viscosity.
It turns out that in a strongly coupled N = 4 super-symmetric Yang Mills theory with a large number of colors 
(the so-called \textquoteright t Hooft limit), 
$\eta$/s can be calculated using a gauge gravity duality and is found to be 1/4$\pi$ (in natural units).
Using the famous AdS/CFT correspondence~\cite{Maldacena:1997re}, 
Kovtun, Son and Starinets conjectured that this value of $\eta$/s is a 
lower bound for all fluids (the KSS bound)~\cite{Kovtun:2004de}. 
We therefore call a fluid with $\eta$/s = 1/4$\pi$ a perfect fluid. 

For many other fluids, like helium, nitrogen, and water, the transition from one phase to another 
occurs in the vicinity where the kinematic viscosity has a minimum. 
Therefore an accurate measurement of the kinematic viscosity as function of the QGP temperature 
would help pinpoint the location of the cross over QCD phase transition~\cite{Csernai:2006zz,Lacey:2006bc}. 

After the first observation of the large elliptic flow at RHIC~\cite{Ackermann:2000tr} it was quickly realised that reported measurements of the elliptic flow included contributions from event-by-event fluctuations~\cite{Miller:2003kd}, 
which were not included in the theoretical description. 
Currently we have strong evidence that these event-by-event fluctuations of 
the anisotropic flow coefficients are mainly due to the event-by-event fluctuations of the initial spatial deformations and 
inhomogeneities of the produced matter~\cite{Miller:2003kd,Andrade:2006yh,Alver:2006wh,Petersen:2008dd,Flensburg:2011wx}. 
The low kinematic viscosity of the QGP makes it possible that traces of these 
initial spatial distributions survive the complete dynamics of the expansion until the final freeze-out stage. 
Therefore, measurements of the event-by-event distribution of the flow coefficients (or the mean values and moments 
of the distribution) and their correlation provide, not only, a unique way to determine the kinematic viscosity of the QGP but also strongly 
constrain the unknown initial conditions.  

The experimental determination of the kinematic viscosity and constraints on the initial conditions can only be achieved 
with precision flow measurements combined with an accurate model description of the 
complete dynamics of the produced system, that is from the initial conditions up to the point 
where the produced particles do not re-interact anymore. 

\section{HEAVY-ION COLLISIONS}
\label{heavy_ions}

Collisions of heavy ions at LHC energies produce a system with temperatures that are well above the strong phase transition temperature 
and are therefore a unique tool to create and study hot and dense QCD matter and its phase transition to ordinary 
hadronic matter in the laboratory.
Like the early universe, the hot and dense system created in a heavy-ion collision will expand and cool down. 
In this time evolution the system probes a range of energy densities and temperatures, and possibly different phases. 
The evolution of the created system can be divided in two characteristic periods. During the formation of the system  
collisions with large momentum transfer occur and during this period the largest energy density is created. 
The system will thermalize and form the QGP provided that the quarks and gluons undergo multiple interactions. 
Due to the thermal pressure, the system undergoes a collective expansion and eventually becomes so dilute that it 
hadronizes. 
In the hadronic phase it further cools down via inelastic and elastic interactions until it becomes non-interacting 
(the freeze- out stage). This collective expansion we call flow.
Therefore, the observation of collective flow signals the presence of multiple interactions between the constituents of the medium produced in the collision. More interactions usually leads to a larger magnitude of the flow and brings the system closer to thermalization. Flow is therefore an observable that provides experimental information on the equation of state and the transport properties of the produced QGP. 

\begin{figure}[htb]
  \begin{center}
    \includegraphics[width=0.99\textwidth]{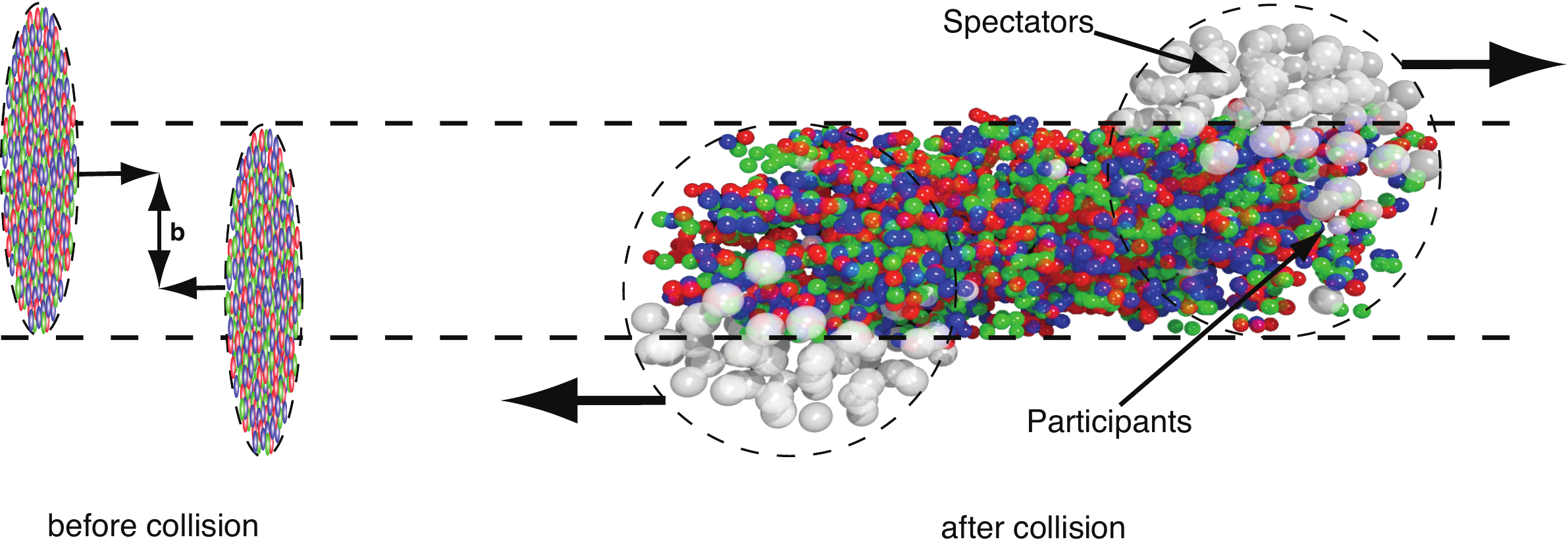}
    \caption{
      Left: The two heavy-ions before collision with impact parameter {\bf b}.
      Right:  The spectators continue unaffected, while in the participant zone particle production takes place.
      \label{centrality} }
  \end{center}
\end{figure}
Heavy ions are extended objects and the system created in a head-on
collision is different from that in a peripheral collision. Therefore,
collisions are categorized by their centrality. 
Theoretically the centrality is 
characterized by the impact parameter ${\bf b}$ (see Fig.~\ref{centrality}) which is, 
however, not a direct observable. 
Experimentally, the collision centrality can be inferred 
from the measured particle multiplicities if one assumes that this
multiplicity is a monotonic function of ${\bf b}$.

In the most central collisions (small impact parameter ${\bf b}$) the spatial distribution of the created system 
is approximately azimuthally symmetric, and the resulting 
azimuthally symmetric expansion is called radial flow. 
For more peripheral collisions the interaction volume changes roughly into an almond shape and due to its spatial anisotropy the expansion 
will be different as function of azimuth. This gives rise to so-called anisotropic flow.

\section{ANISOTROPIC FLOW}
\label{collective_flow}

Experimentally, the most direct evidence of collective flow comes from the observation 
of anisotropic flow, which is the anisotropy in particle momentum distributions correlated 
with the flow symmetry plane. 
The various patterns of anisotropic
flow are characterised using a Fourier expansion of the event averaged azimuthal particle distribution:
\begin{equation}
  \frac{dN}{d\varphi} = \frac{\bar{N}}{2\pi} \left(1 + 2 \sum_{n=1}^{\infty} \bar{v}_n \cos(n(\varphi{-}\bar{\Psi}_n)) \right),
  \label{invariantyield}
\end{equation}
where $\bar{N}{\,\equiv\,}\langle N\rangle$ is the mean number of selected particles per event, $\varphi$ the azimuthal angle, 
and $\bar{\Psi}_n$ the mean angle of the $n$-th harmonic flow plane. 

The Fourier coefficients are in general $\pt$ and rapidity ($y$) dependent and are given by:
\begin{eqnarray}
\bar{v}_n(\pt,y) &=& \langle \langle \cos[n(\varphi{-}\bar{\Psi}_n)] \rangle \rangle
\;\;\; {\rm or\ equivalently} \nonumber \\ 
\bar{v}_n(\pt,y)  &=& \langle \langle e^{in\varphi}  e^{-in\bar{\Psi}_n} \rangle \rangle,
\label{fouriercoeff}
\end{eqnarray}
where $\langle \langle \dots \rangle \rangle$ denotes an average in the $(\pt,y)$ bin under study over particles within 
the same event, and is used in the context of single-particle, two-particle and multi-particle averages, 
followed by averaging over all events.  
Since the Fourier series in Eq.~\ref{fouriercoeff} is expressed only in terms of cosines, 
the imaginary part of the second line is zero. 
In this Fourier decomposition, the coefficients $\bar{v}_1$ and $\bar{v}_2$ 
are known as directed and elliptic flow, respectively.

\subsection{Experimental Methods}
\label{methods}

Because the flow angle is not a direct observable the anisotropic flow (Eq.~\ref{fouriercoeff}) can not be measured directly,
therefore it is usually estimated using azimuthal correlations between the observed particles. 
Two-particle azimuthal correlations, for example, can be written as:
\begin{eqnarray}
\langle\langle e^{in(\varphi_1 - \varphi_2)} \rangle\rangle &=& \langle\langle e^{in(\varphi_1 - \bar{\Psi}_n - 
(\varphi_2 - \bar{\Psi}_n))} \rangle\rangle , 
\nonumber\\
&=& \langle\langle e^{in(\varphi_1 - \bar{\Psi}_n)} \rangle \langle e^{-in(\varphi_2 - \bar{\Psi}_n)} \rangle + \delta_{2,n}\rangle, 
\nonumber\\
&=& \langle v_n^2 + \delta_{2,n}\rangle, 
\label{twoParticleFlowEstimate3}
\end{eqnarray}
In Eq.~\ref{twoParticleFlowEstimate3} we have factorized the azimuthal correlation 
between the particles in a common correlation with the flow angle (anisotropic flow $v_n$) and a correlation 
independent of the flow angle (nonflow $\delta_{2,n}$). 
If $\delta_{2,n}$ is small, Eq.~\ref{twoParticleFlowEstimate3} can be used to measure $\left<v_{n}^{2}\right>$, but in general the nonflow contribution is not negligible. These additional nonflow correlations arise from e.g. resonance decays, jet fragmentation, and Bose-Einstein correlations. 
They can be suppressed by appropriate kinematic cuts (which complicate comparisons between experiments) 
or by making use of the collective nature of anisotropic flow using multi-particle correlations. 
 
For the latter, this is done using so called cumulants~\cite{Borghini:2001vi}, which are genuine multi-particle correlations. 
For instance, the two-particle cumulant $c_{n}\{2\}$ and the four-particle cumulants $c_{n}\{4\}$ are defined as:
\begin{eqnarray}
c_{n}\{2\} &\equiv& \left<\left<e^{in(\varphi_1-\varphi_2)}\right>\right> = \left<v_n^2 + \delta_{2,n}\right>,
\label{twoParticleFlowEstimate4}\\
c_{n}\{4\} &\equiv& \left<\left<e^{in(\varphi_{1} + \varphi_{2}-\varphi_{3}-\varphi_{4})}\right>\right>-2\left<\left<e^{in(\varphi_1-\varphi_2)}\right>\right>^{2},
\nonumber\\
&=& \left<v_{n}^{4} + \delta_{4,n} + 4 v_{n}^{2}\delta_{2,n} + 2\delta_{2,n}^{2}\right> - 2\left< v_{n}^{2}+ \delta_{2,n}\right>^{2},  
\nonumber\\
&=& \left<-v_{n}^{4} + \delta_{4,n}\right>.
\label{fourParticleFlowEstimate3}
\end{eqnarray}
From the combinatorics it is easy to show that 
$\delta_{2,n} \propto 1/M_{\rm c}$ and $\delta_{4,n} \propto 1/M_{\rm c}^{3}$, where $M_{\rm c}$ is the number of independent particle clusters.
Therefore, $v_{n}\{2\}$ is only a good estimate if $v_{n} \gg 1/\sqrt{M_{\rm c}}$ while  $v_{n}\{4\}$ is already a good estimate of 
$v_{n}$ if $v_{n} \gg 1/{M_{\rm c}}^{3/4}$; for $c_n\{\infty\}$ this argument leads to $v_{n} \gg 1/M_{\rm c}$. 
This shows that for a typical Pb--Pb collision at the LHC with $M_{\rm c} = 500$ the possible nonflow contribution can be reduced by more than an order of magnitude using higher order cumulants.  

One of the problems in using multi-particle correlations is the computing power needed to go over all possible particle multiplets. 
To avoid this problem, multi-particle azimuthal correlations in heavy-ion collision are often calculated using so-called $Q$-cumulants~\cite{Bilandzic:2010jr,Bilandzic:2013kga}. 
The $Q$-cumulants are calculated  analytically from a flow vector $Q_n$: 
\begin{equation}
Q_{n} \equiv \sum_{i=1}^M 
e^{in\varphi_i}\,,
\label{Qvector}
\end{equation}
where $M$ is the number of selected particles in an event. In general this flow vector can be constructed using weights for the individual particles and the angle of this flow vector $Q_n$ can be used to estimate the 
flow angle, similar to the first introduction of $Q_n$ in~\cite{Danielewicz:1985hn}. 
As an example we show how the flow vector is used to calculate the two-particle cumulant.

Realising that 
\begin{equation}
\left|Q_n\right|^2 = \sum_{i,j=1}^{M} e^{in(\varphi_i-\varphi_j)}\,
= M+\sum_{i \ne j} e^{in(\varphi_i-\varphi_j)}\,,
\label{|Q_n|^2}
\end{equation}
and using this in our definition of the individual event averaged two-particle azimuthal correlation 
\begin{equation}
\left< e^{in(\varphi_{1} - \varphi_{2})} \right> \equiv \frac{1}{M(M -1)}\,
\sum_{i \ne j} e^{in(\varphi_i-\varphi_j)} =
\frac{\left|Q_n\right|^2 -M}{M(M -1)} \,,
\label{2pCorrelationSingleEvent}
\end{equation}
we immediately see that we can express the individual event average two-particle azimuthal correlation 
in terms of powers of the flow vector.
With the two-particle cumulant defined as the average over all events of Eq~\ref{2pCorrelationSingleEvent} we get: 
\begin{eqnarray}
c_{n}\{2\}  
&\equiv&  
\left<\!\left< e^{in(\varphi_{1} - \varphi_{2})} \right>\!\right>  
 \equiv
\frac{\ds \sum_{i=1}^{N} M_i(M_i-1) \left< e^{in(\varphi_{1} - \varphi_{2})} \right>_i}
{\ds \sum_{i=1}^{N}
  M_i(M_i-1)} \nonumber \\
&=&
\frac{  \ds \sum_{i=1}^{N} \left|Q_{in}\right|^2 - M_i}
{\ds \sum_{i=1}^{N}
  M_i(M_i-1)}  
  \,,
\label{2pCorrelationAllEvents}
\end{eqnarray}
%
where $N$ is the number of events.
This expression already simplifies the calculation of the two-particle cumulant because it does not require a nested 
loop over all the particles pairs.
For four-particle and higher order cumulants the equations can be found in~\cite{Bilandzic:2010jr,Bilandzic:2013kga} 
and for these calculations the gain from using powers of the $Q$-vector are even bigger.  
 
To relate the two- and multi-particle observables to the flow Fourier coefficients in Eq.~\ref{fouriercoeff} we 
already mentioned that we assumed that the correlation between $v_n$ and $\delta_{2,n}$ is negligible. 
In addition the assumption was made that $\langle \delta_{2,n}^{2} \rangle = \langle \delta_{2,n} \rangle^2$ and 
$\langle v_{n}^{4} \rangle = \langle v_{n}^{2} \rangle^2$.
In other words, we have neglected the event-by-event fluctuations in $v_n$ and $\delta_{2,n}$.
The effect of the fluctuations on $v_n$ estimates can be obtained from
\begin{eqnarray}
    \langle v_n^2 \rangle &=& \bar{v}_n^2 + \sigma^2_{v_n}, \nonumber \\
    \langle v_n^4 \rangle &=& \bar{v}_n^4 + 6\sigma^2_{v_n} \bar{v}_n^2, \nonumber \\
    \langle v_n^6 \rangle &=& \bar{v}_n^6 + 15\sigma^2_{v_n} \bar{v}_n^4, 
    \label{sigma} 
\end{eqnarray}
where $\sigma_{v_n}$ is the variance of $v_n$.
Neglecting the nonflow terms we have the following expressions for the cumulants:
\begin{eqnarray}
    v_n\{2\} &=& \sqrt{\langle v_n^2 \rangle}, \nonumber \\
    v_n\{4\} &=& \sqrt[4]{2 \langle v_n^2\rangle^2 - \langle v_n^4 \rangle}, \nonumber \\
    v_n\{6\} &=& \sqrt[6]{\frac{1}{4} \left( \langle v_n^6 \rangle -9 \langle v_n^2\rangle \langle v_n^4\rangle + 12 \langle v_n^2\rangle^3 \right)}. 
    \label{fcummu}
\end{eqnarray}
Here we have introduced the notation $v_n\{k\}$ as the flow estimate from the cumulant $c_n\{k\}$.
In case that $\sigma_{v_n} \ll \bar{v}_n $ we obtain from Eqs.~\ref{sigma} and~\ref{fcummu}, up to order $\sigma^2_{v_n}$:
\begin{eqnarray}
    v_n\{2\} &=& \bar{v}_n + \frac{1}{2} \frac{\sigma^2_{v_n}}{\bar{v}_n} ,\nonumber \\
    v_n\{4\} &=& \bar{v}_n  - \frac{1}{2} \frac{\sigma^2_{v_n}}{\bar{v}_n} ,\nonumber \\ 
    v_n\{6\} &=&  \bar{v}_n - \frac{1}{2} \frac{\sigma^2_{v_n}}{\bar{v}_n}.
    \label{fluctuations1}
\end{eqnarray}
Equations~\ref{fourParticleFlowEstimate3} and~\ref{fluctuations1} illustrate that the difference between 
$v_{n}\{2\}$ and $v_{n}\{4\}$ is sensitive to not only nonflow but also to the event-by-event $v_n$ fluctuations.

\subsection{Initial Conditions}
\label{inititial_conditions}
The development of anisotropic flow is controlled by the anisotropies in the pressure gradients which in turn depend 
on the shape and structure of the initial density profile. 
The latter can be characterised, in analogy with the flow Fourier coefficients and flow angles of Eq.~\ref{fouriercoeff},  
by a set of harmonic eccentricity coefficients $\varepsilon_n$ and associated angles $\Phi_n$:
%
\begin{eqnarray}
\label{eq3.1}
\varepsilon_1 e^{i\Phi_1} &\equiv& - \frac{\int  r \, dr \, d\phi \, r^3 e^{i\phi} \, e(r,\phi)}
                                                            {\int r \, dr \, d\phi \, r^3 e(r,\phi)}, \nonumber  \\
\varepsilon_n e^{in\Phi_n} &\equiv& - \frac{\int  r \, dr \, d\phi \, r^n e^{in\phi} \, e(r,\phi)}
                                                              {\int r \, dr \, d\phi \, r^n e(r,\phi)}\  (n>1),
\end{eqnarray}
%
where $e(r,\phi)$ is the initial energy density distribution in the plane transverse to the beam direction.
It should be noted, however, that these eccentricity coefficients involve only the lowest moments 
of the initial density profile, describing the large scale structures relevant for the hydrodynamic response.

Even at a fixed impact parameter, due to event-by-event fluctuations of the transverse positions of the nucleons inside the colliding nuclei~\cite{Miller:2003kd}, and of the gluon density profiles inside those nucleons~\cite{Kovner:1995ja,Kovchegov:1997ke,Krasnitz:1998ns,Lappi:2003bi,Schenke:2012wb,Muller:2011bb,Dumitru:2012yr} the $\varepsilon_n$ and $\Phi_n$ vary event-by-event. 
The event-by-event $\varepsilon_n$ and $\Phi_n$ control in a hydrodynamic picture the event-by-event 
anisotropic flow coefficients $v_n$ and their directions $\Psi_n$. 
The different cumulants are sensitive to the statistical distribution of these flow coefficients and angles, and can be used to constrain the current uncertainty on the initial conditions.
  
\begin{figure}[htb]
  \begin{center}
    \includegraphics[width=0.5\textwidth]{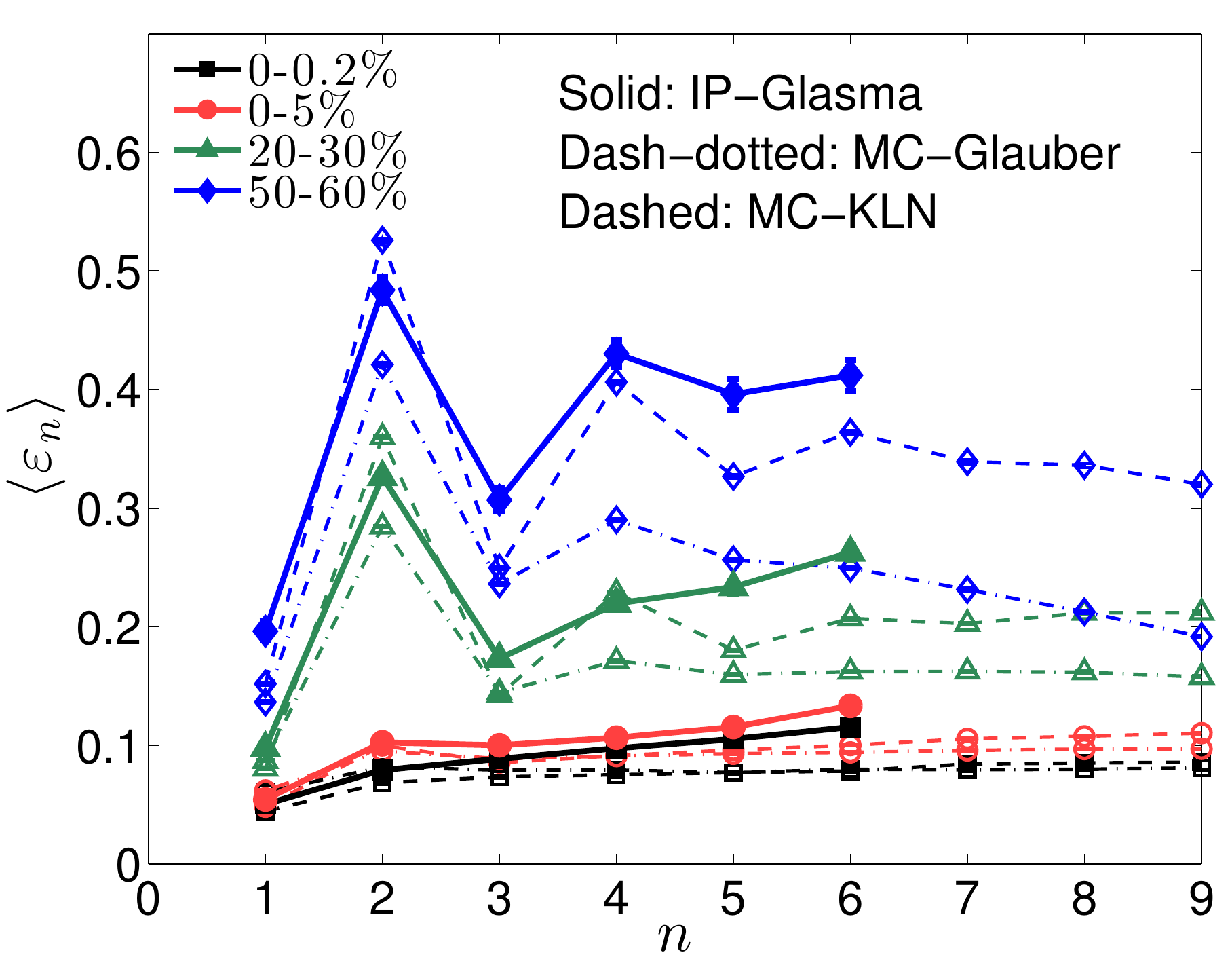}
    \caption{
      The $\varepsilon_n$ rms from three different initial-state models (IP-Glasma, MC-Glauber, MC-KLN) for 
      2.76\,$A$\,TeV Pb+Pb collisions of different centralities (figure from~\cite{Heinz:2013th}).
      \label{eccentricity} }
  \end{center}
\end{figure}
This is important because in heavy-ion collisions the physics of the initial stage is one of the biggest open questions, and is responsible for the largest uncertainty in many observables. 
Currently different theoretical models are used to model the initial energy and entropy density. Figure~\ref{eccentricity} shows the corresponding 
$\langle \varepsilon_n \rangle$ distribution for the Monte Carlo (MC) Glauber~\cite{Miller:2007ri}, 
Monte Carlo Kharzeev-Levin-Nardi (MC-KLN)~\cite{Drescher:2006ca}  
and IP-Glasma models~\cite{Bartels:2002cj,Moreland:2012qw,Kovner:1995ja,Kovchegov:1997ke,KL2012,Krasnitz:1998ns,Lappi:2003bi}, 
for four centrality classes.
The MC-Glauber is the most common model used. 

In the MC-Glauber model the entropy is proportional to the number of interacting nucleons in the 
two colliding nuclei. The fluctuating event-by-event distribution of these nucleons are in the MC-Glauber 
model responsible for the fluctuations in $\varepsilon_n$. 
In the MC-KLN model the entropy is calculated by the initial gluon production, 
which is position dependent and is calculated based on Color Glass Condensate ideas. 
Both these models do not take into account fluctuations of the gluon fields inside the colliding nucleons. 
The IP-Glasma model does take this into account and in addition evolves these gluon fields, 
using classical Yang-Mills dynamics, to a matching surface after which hydrodynamic model calculations can take over. 

In Fig.~\ref{eccentricity} the $\langle \varepsilon_n \rangle$ are plotted for these three models in different centrality bins.
In the most central collisions, 0--0.2\%, the $\varepsilon_n$ are entirely due to fluctuations, and 
all $\langle \varepsilon_n \rangle$ have roughly equal magnitudes.
For increasingly less central collisions the spatial distribution of the produced system becomes more almond like which is clearly seen by the increase of $\langle \varepsilon_2 \rangle$ and also, although to a lesser extend, in $\langle \varepsilon_4 \rangle$ and $\langle \varepsilon_6 \rangle$. The odd eccentricity coefficients are on the other hand for all centralities dominated 
by fluctuations. 

There is a clear difference between the three models for the full $\langle \varepsilon_n \rangle$ spectrum and 
measurements of the anisotropic flow coefficients allow us, in principle, to discriminate between these various 
models of the initial state. 
This is because the $\langle v_n \rangle$ depend on the anisotropies in the pressure gradients which in turn 
depend on $\langle \varepsilon_n \rangle$. 
In addition, the theoretical models also differ in their predictions for the correlations between the eccentricity 
coefficients $\langle \varepsilon_n \rangle$ (not shown here) which again can be tested experimentally with the
correlations between the different $\langle v_n \rangle$. 
These correlations between eccentricity coefficients $\langle \varepsilon_n \rangle$ are modified through a 
non-linear viscous hydrodynamics evolution~\cite{Gardim:2011xv,Teaney:2012ke,Qiu:2012uy}, which provides even more interesting tests of the 
hydrodynamical paradigm.  

\subsection{Viscous effects on anisotropic flow}
\label{viscous_effects}
Lattice QCD calculations of the equation of state of a QGP showed only small deviations from the Stefan-Boltzmann 
limit which matched the expectation that the QGP would behave like an massless ideal gas. 
However, unexpectedly, the experimental data 
on elliptic flow from RHIC showed good agreement with model predictions from ideal hydrodynamics. 
This dramatically changed our understanding of the QGP, which resembles, instead of a dilute gas of quarks and gluons, a rather strongly coupled liquid.

While ideal hydrodynamics was very successful, it was only applicable for relatively low transverse momenta 
and more central collisions. 
The realisation that small viscous corrections significantly affect the buildup of 
anisotropic flow~\cite{Teaney:2003kp,Romatschke:2007mq,Luzum:2008cw} triggered 
the theoretical development of viscous hydrodynamical models~\cite{Romatschke:2007mq,Luzum:2008cw,Baier:2007ix,Muronga:2003ta,Heinz:2005bw,Song:2007fn,Song:2008si,Molnar:2009tx,Denicol:2010xn,Denicol:2012cn,El:2009vj}. 
It also triggered the comparison with models of strongly interacting systems such as the  
AdS/CFT correspondence~\cite{Maldacena:1997re}. 
These calculations showed that the kinematic viscosity $\eta/s$ should be very small~\cite{Kovtun:2004de} to explain the observed large elliptic flow and that the 
EoS of the QGP indeed changes very little from an ideal gas to a strongly coupled system~\cite{Gubser:1996de}.  

The shear viscosity reduces the difference  between the expansion velocities in the system and therefore reduces the anisotropic flow. This reduction is stronger for larger $n$ and, as a consequence, the spectrum of $v_n$ 
coefficients is a very sensitive observable to determine the magnitude of $\eta/s$. 
It is thought that viscous corrections increase for particles with larger transverse momenta. 
Therefore the measurements of $v_n$ at high 
transverse momenta are in principle a sensitive probe of  $\eta/s$.
In ideal hydrodynamics switching from a fluid description to a distribution of particles $f(x,p)$ is understood using 
the Cooper-Frye prescription. 
For a system which is slightly out of equilibrium, i.e. a viscous system, this process is not so well understood and is 
modelled with a small deviation from the equilibrium distribution adding a term $\delta f$.
Because $\delta f$ increases roughly as $\pt^{\alpha}$ also the theoretical uncertainties increase 
with increasing $\pt$, therefore the comparison between the $v_n$ coefficients and hydrodynamics are most reliable 
at small transverse momenta, typically $\pt \leq 2$ GeV/$c$.  

\section{ANISOTROPIC FLOW MEASUREMENTS FROM ALICE}
\label{elliptic_flow}
While the large elliptic flow observed at RHIC provided compelling evidence 
for strongly interacting matter, a precise determination of $\eta/s$ in the QGP was complicated by uncertainties in 
the initial conditions of the collision, the relative contributions to the anisotropic flow from the hadronic and partonic phase, 
and the unknown temperature dependence of $\eta/s$. 
Because of all these uncertainties it was not even clear if the elliptic flow would increase 
or decrease when going from RHIC to LHC energies; the first measurement of elliptic flow at the LHC was therefore 
one of the most anticipated results.

\begin{figure}[h!]
\includegraphics[width=0.47\textwidth]{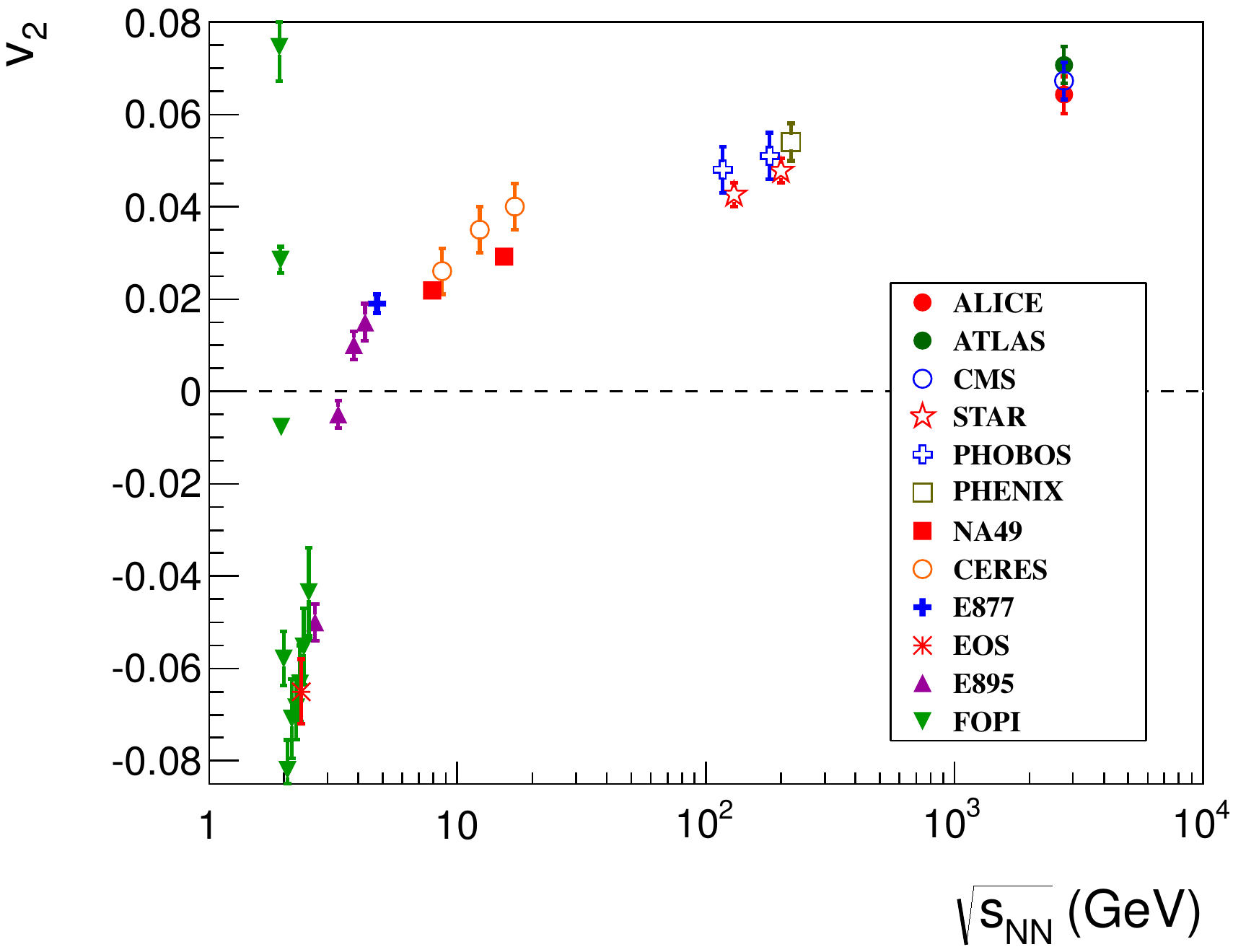}
\includegraphics[width=0.53\textwidth]{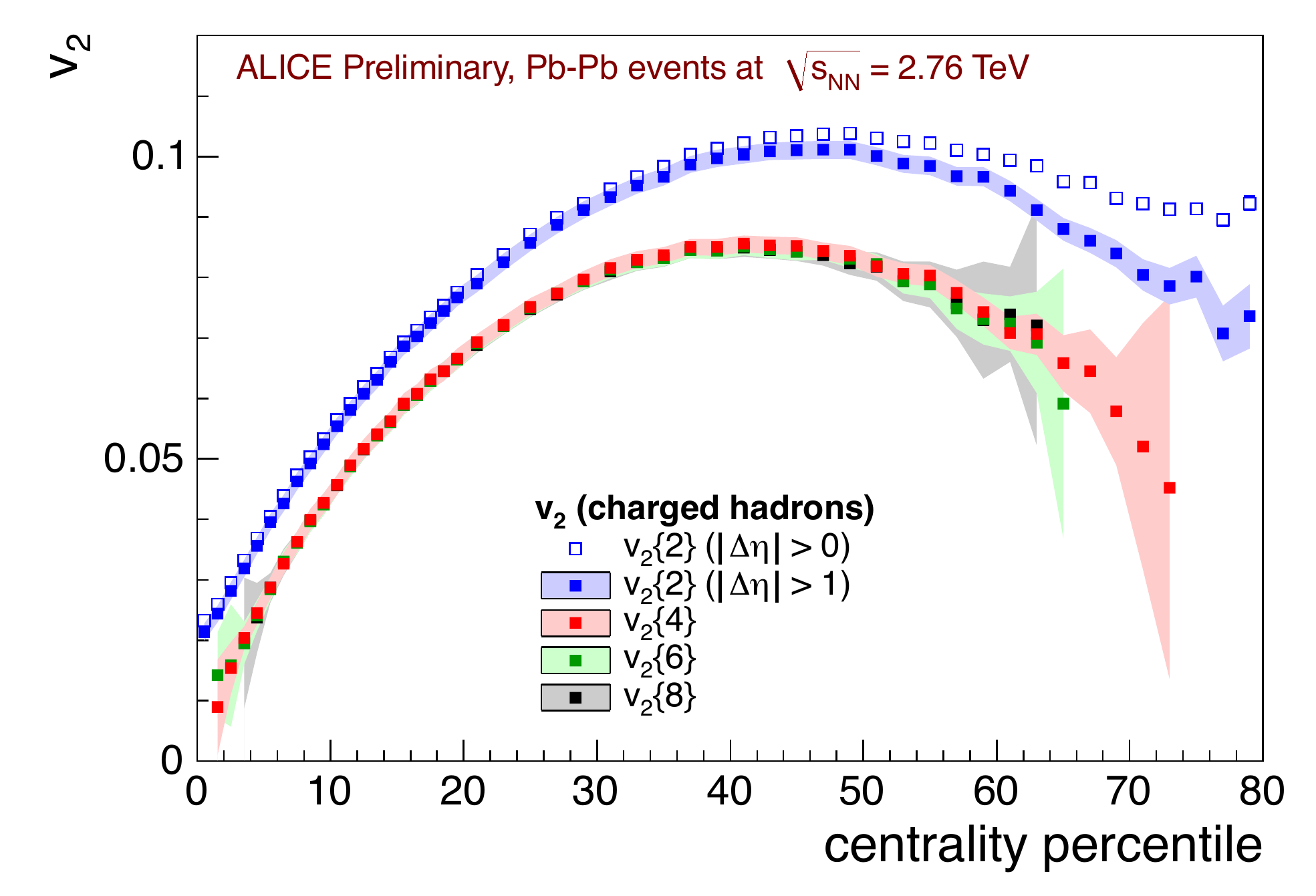}
\caption{
Integrated elliptic flow as a function of collision energy for the 20--30\% centrality class 
(left)~\cite{Aamodt:2010pa,ATLAS:2011ah,Chatrchyan:2012ta} and (right) integrated elliptic flow estimates as a function of collision centrality (right figure from~\cite{Bilandzic:2011ww}).  
}
\label{fig:integratedv2} 
\end{figure}

In the left panel of Fig.~\ref{fig:integratedv2} the LHC measurements at 
2.76 TeV~\cite{Aamodt:2010pa,ATLAS:2011ah,Chatrchyan:2012ta} show that the integrated elliptic flow 
of charged particles increases by about 30\% compared to the elliptic flow 
measured at the highest RHIC energy of 0.2 TeV. 
This result indicates that the hot and dense matter created in these collisions at the LHC still behaves like a fluid with almost 
zero friction, and in addition, provides strong constraints on the temperature dependence of $\eta/s$.

The centrality dependence, in narrow bins (1--2\%) to reduce trivial event-by-event fluctuations, of the elliptic flow at the 
LHC is plotted in the right panel of Fig.~\ref{fig:integratedv2}.
The elliptic flow shows an expected 
increase with decreasing centrality because of the increasing initial spatial anisotropy of the collision zone. 
For more peripheral collisions the elliptic flow starts to decrease again, which is attributed mainly due to 
larger viscous corrections in the smaller system with larger gradients. 

There is a significant difference between flow estimates from two- and multi-particle correlations.
This difference is caused by nonflow contributions and by event-by-event fluctuations in the elliptic flow.
The effect of the nonflow on two-particle estimates is apparent 
in more peripheral collisions from the difference in $v_2$ calculated from the correlation 
between particles with a gap in pseudorapidity $|\Delta\eta| > 0$ and $|\Delta\eta| > 1$. 
The results from four-, six-, and eight-particle cumulant estimates which are consistent within uncertainties, 
indicate that the genuine four-particle, and higher order, nonflow contribution is negligible. The contribution 
of the event-by-event fluctuations is discussed in the next section.

\begin{figure}[h!]
\includegraphics[width=0.54\textwidth]{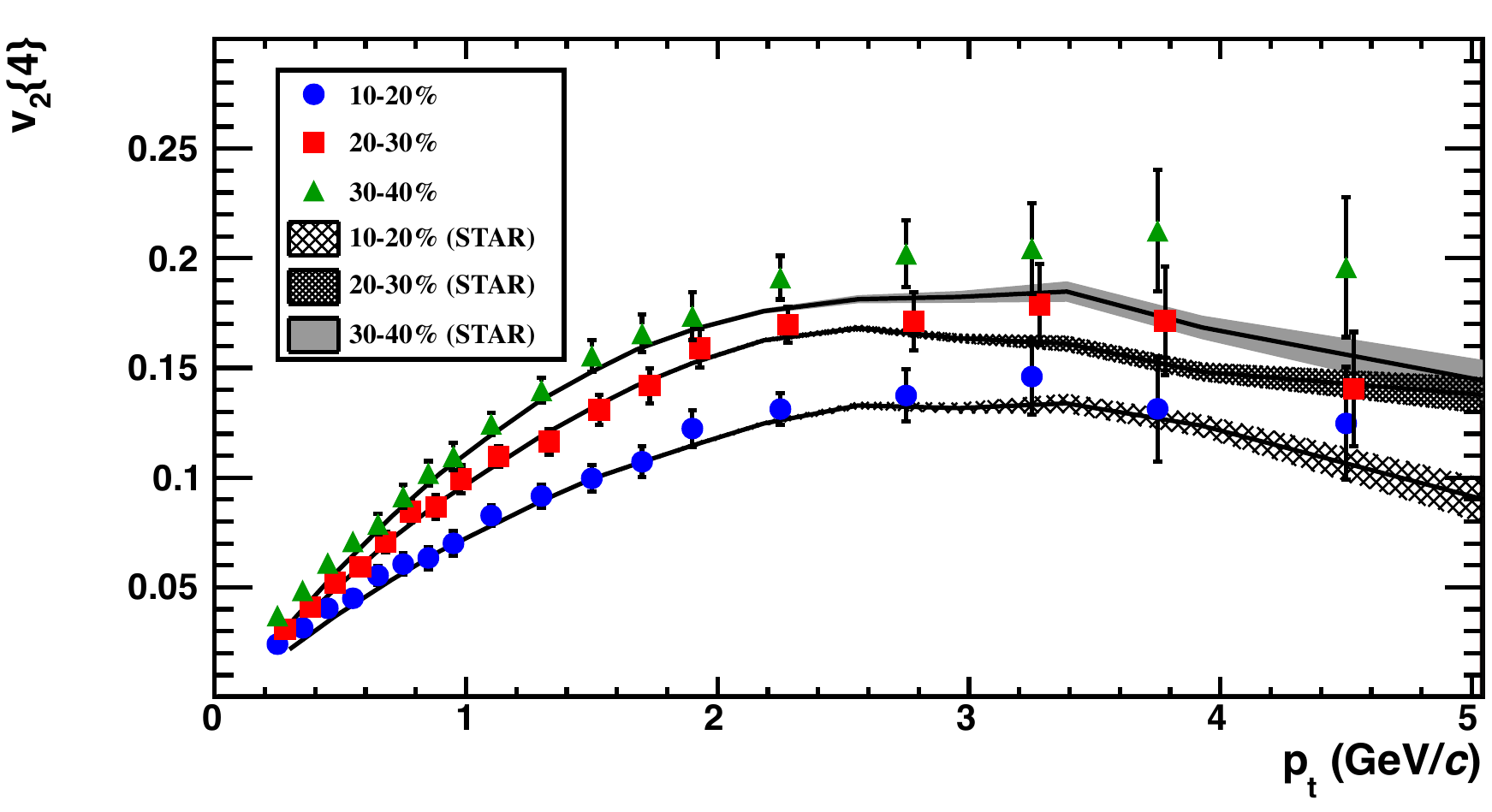}
\includegraphics[width=0.46\textwidth]{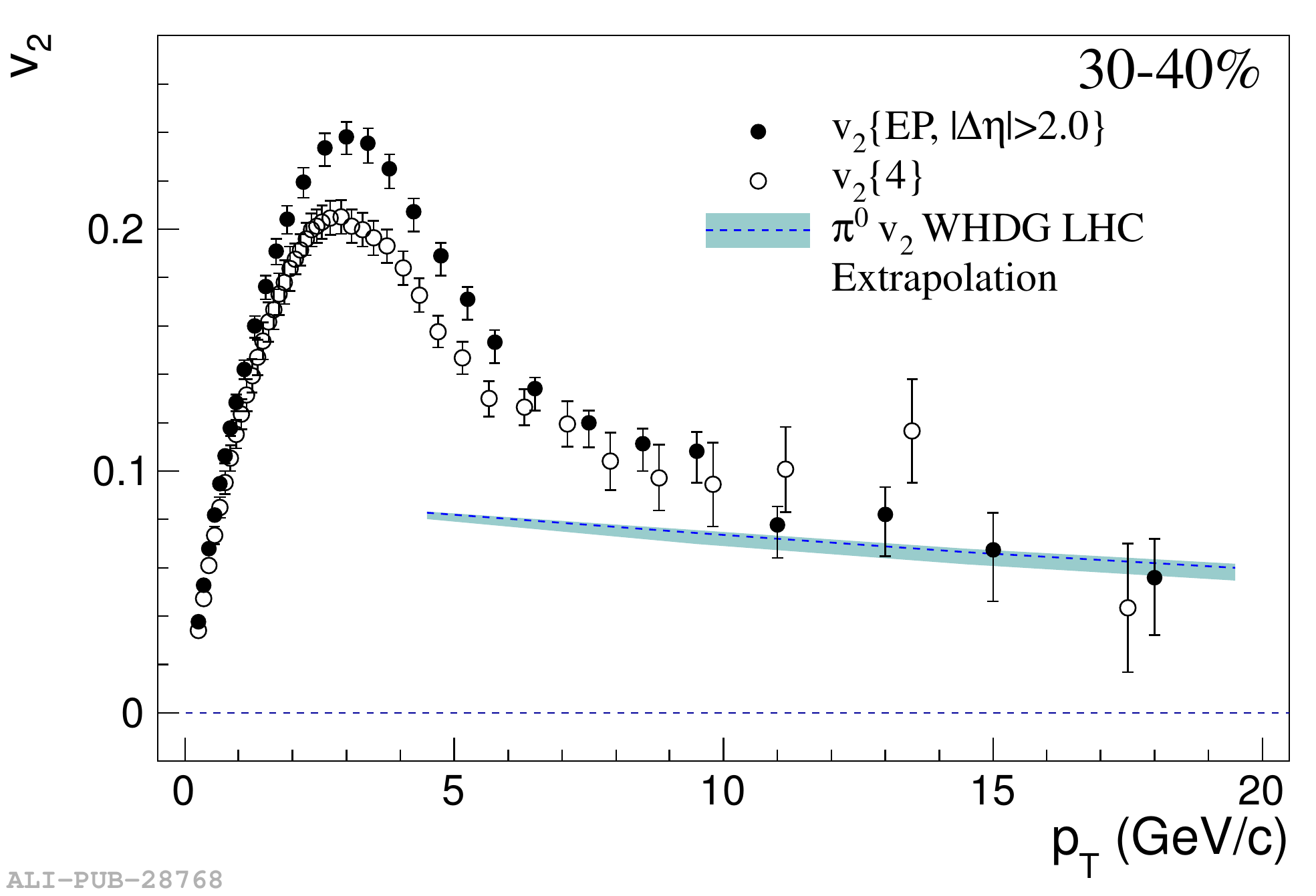}
\caption{
Left: A comparison between $\pt$-differential elliptic flow of charged particles as a function of centrality at RHIC and LHC 
energies (figure from~\cite{Aamodt:2010pa}).
Right: $\pt$-differential elliptic flow of charged particles compared to jet energy loss model calculations 
(figure from~\cite{Abelev:2012di}).
}
\label{fig:v2pt} 
\end{figure}
Elliptic flow as a function of transverse momentum is, compared to integrated flow, more sensitive to 
the evolution and freeze-out conditions of the produced system. 
The left panel of Fig.~\ref{fig:v2pt} shows the charged particle $\pt$-differential $v_2\{4\}$ measured 
in ALICE compared to STAR measurements at RHIC. 
Remarkably, the $v_2(\pt)$ is the same within uncertainties at low $\pt$~\cite{Aamodt:2010pa}, 
while the beam energies are different by more than one order of magnitude. 

The $\pt$-differential $v_2$, plotted in the right panel of Fig.~\ref{fig:v2pt}, 
is still large above 15 GeV/$c$. At these relatively high transverse momenta the hadron yields 
are not from a boosted thermal system but are thought to contain a dominant contribution from the fragmentation of high energy partons produced at initial hard scatterings. 
These high energy partons traversing the nuclear matter are expected to lose energy, which depends strongly on the color charge density of the matter and on the path length traversed by the partons.
Because the spatial shape of the created system is anisotropic this path length depends on the 
azimuthal emission angle, 
which introduces an azimuthal anisotropy in the particle emission at large $\pt$~\cite{Snellings:1999gq}. 
The magnitude of the observed $v_2$ at high-$\pt$ is roughly consistent with model calculations of 
the effect of parton energy loss for neutral pion production~\cite{Horowitz:2011cv}.

\begin{figure}[h!]
\includegraphics[width=0.5\textwidth]{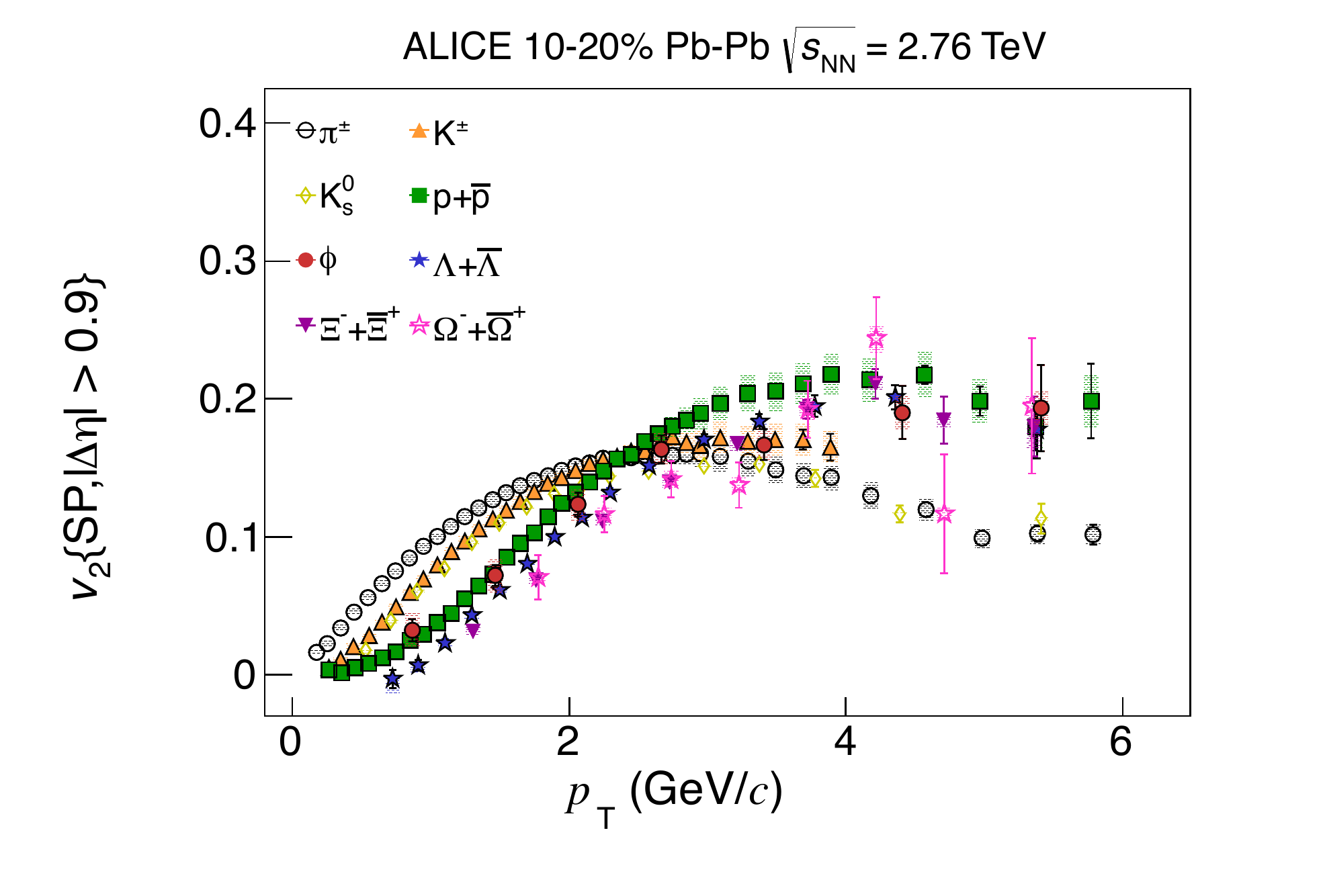}
\includegraphics[width=0.5\textwidth]{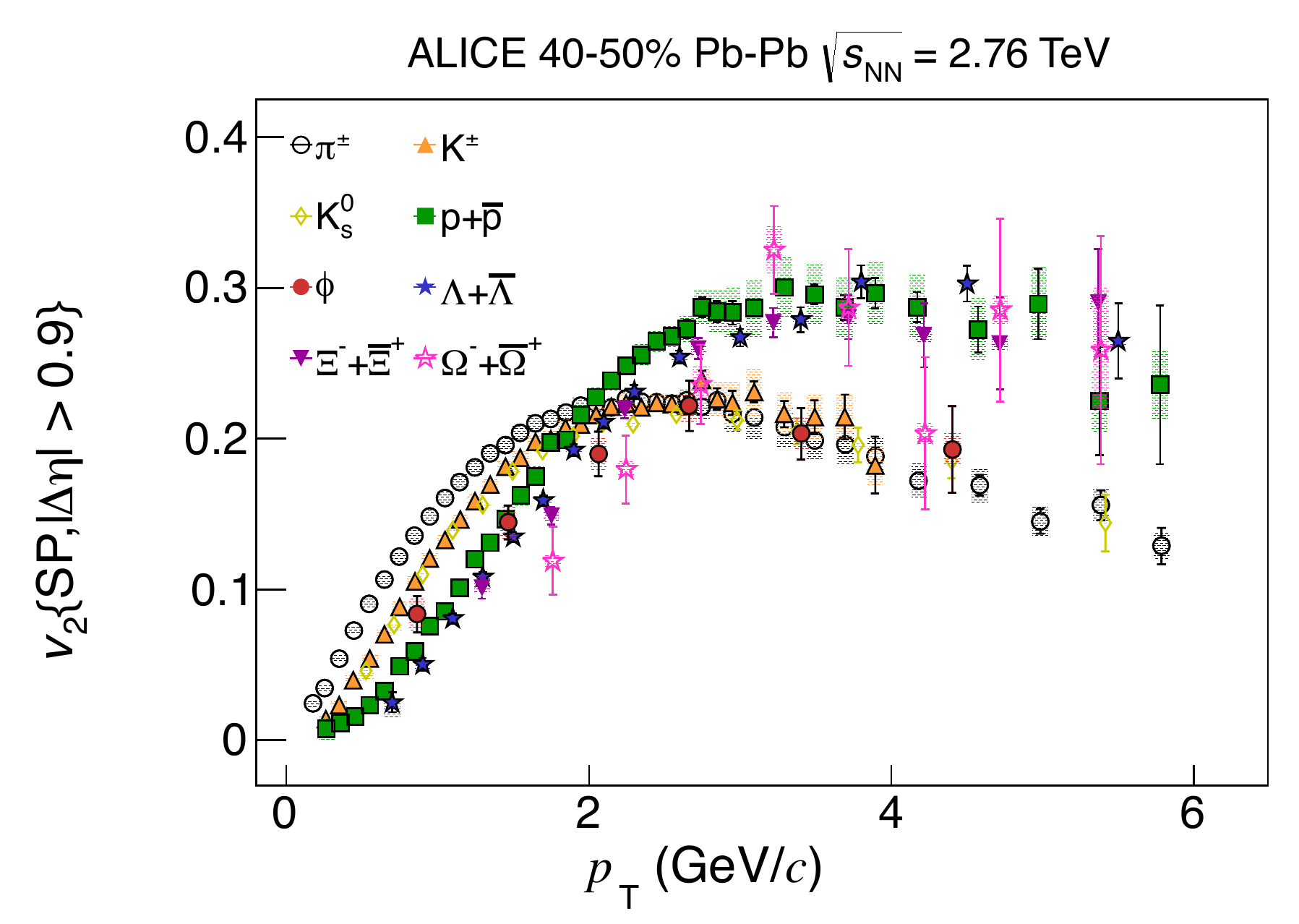}
\caption{
$\pt$-differential elliptic flow of identified particles for the 10--20\% (left) and 40--50\% (right) centrality class
(figures from~\cite{Abelev:2014pua}).
}
\label{fig:identifiedv2} 
\end{figure}The 30\% increase in the integrated flow, which was shown in the left panel of Fig.~\ref{fig:integratedv2}, 
is clearly not due to the change of $v_2(\pt)$ as function of beam energy and must therefore be due to 
an increase in average transverse momentum~\cite{Abelev:2013vea}.
This increase in $\langle \pt \rangle$ is in a hydrodynamical picture naturally explained by an increase 
in the radial flow. This modifies the spectra with a blueshift which depends on the mass of the particle 
and leads to flatter $\pt$ distributions, particular at low-$\pt$.
In hydrodynamics this blueshift is stronger in the direction of the flow plane which results in a depletion 
of low-$\pt$ particles in the direction of this plane. 
This depletion becomes stronger for increasing particle mass and radial flow, 
which is the reason that at a fixed value of $\pt$ heavier particles have a 
smaller $v_2$ (this is also true for other $v_n$ coefficients) compared to lighter ones~\cite{Huovinen:2001cy}. 
Figure~\ref{fig:identifiedv2} shows the $v_2$ for particles of different mass 
(with masses which differ by an order of magnitude) for two centralities.
A mass ordering is observed, as was also seen in STAR~\cite{Adler:2001nb}, 
in both centrality bins, left panel 10-20\% and right 40-50\%, 
and for all particles. This mass ordering is broken at intermediate $\pt$ (2-3 GeV/$c$) which is not expected 
in ideal hydrodynamics. 

The $\phi$-meson is among the particle species of special interest as its mass is very close to the proton and 
$\Lambda$ baryons. Below $\pt = 2.5$ GeV/$c$ the $\phi$-meson follows within the relatively large uncertainties the mass hierarchy for all centralities. 
However, for the lowest $\pt$ bin there is an indication that the $\phi$-meson $v_2$ is larger than the proton $v_2$.
\begin{figure}[h!]
\includegraphics[width=0.5\textwidth]{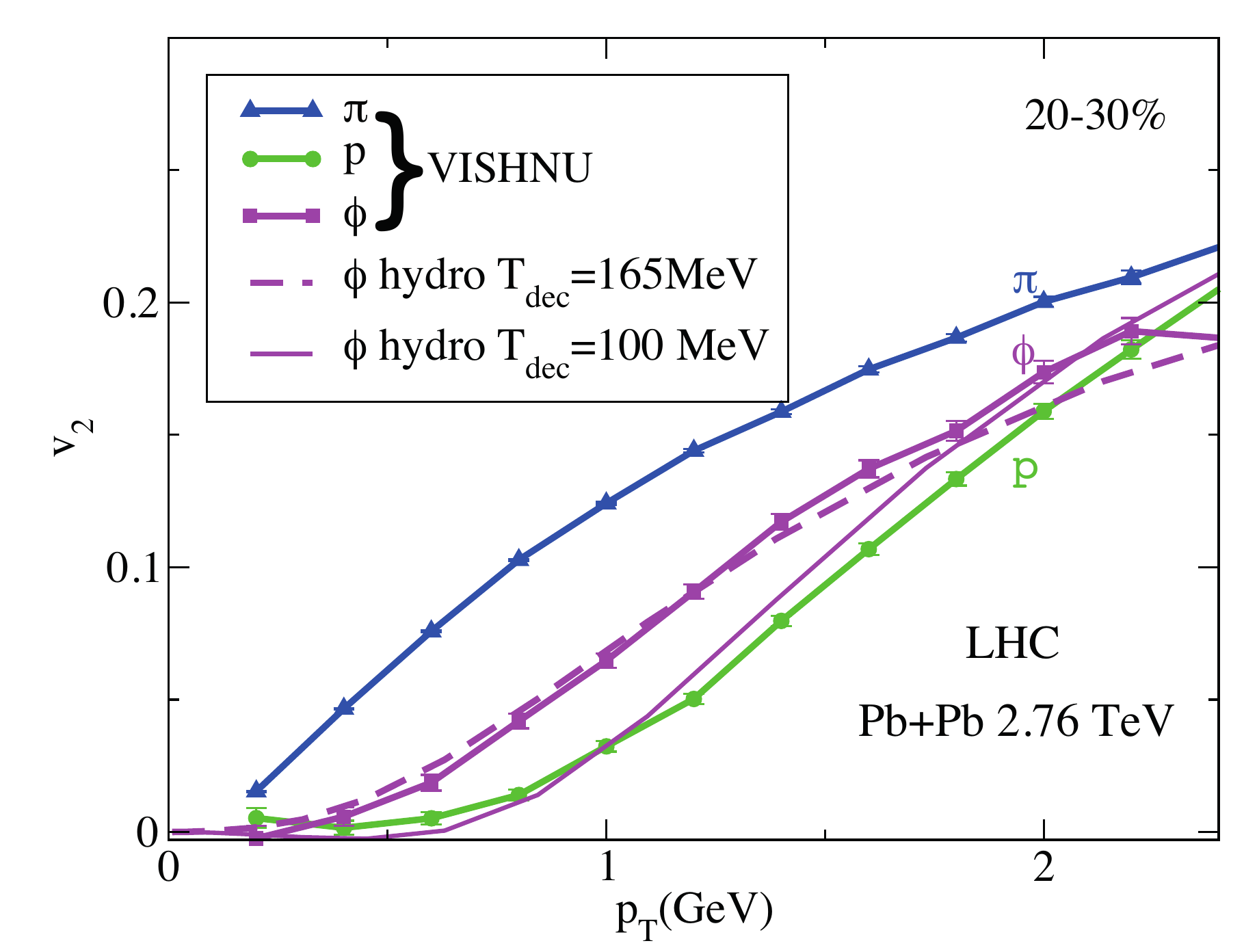}
\includegraphics[width=0.5\textwidth]{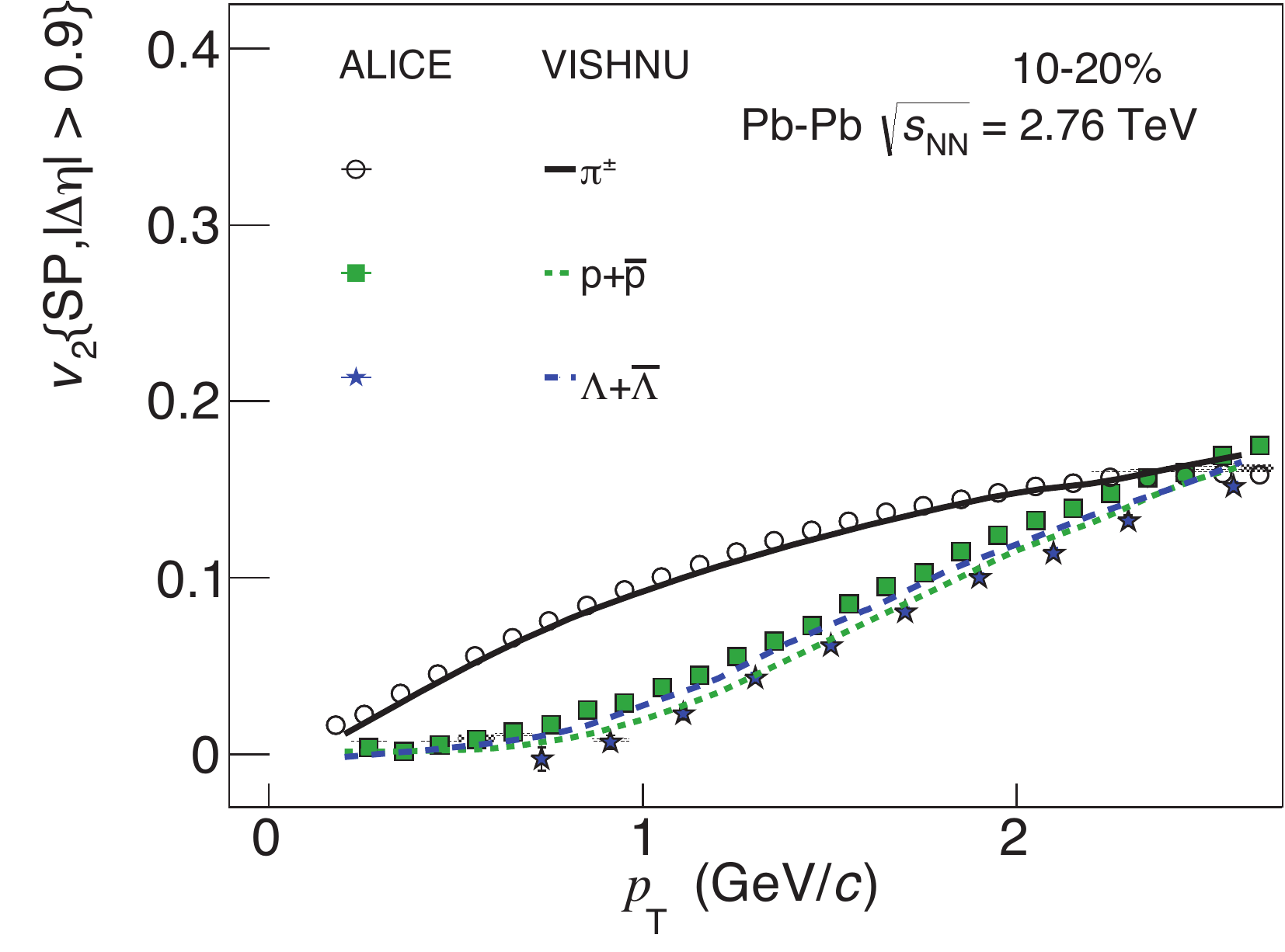}
\caption{
Left: A comparison between {\sc VISHNU} hybrid model calculations, that couple viscous hydrodynamical calculations 
to a hadronic transport model, for $\pi$, p and $\phi$ $v_2$ and a full hydro calculations for the $\phi$ $v_2$ 
with two decoupling temperatures $T_{\rm dec}$ (figure from~\cite{Song:2013qma}). 
Right: A comparison between ALICE measurements and {\sc VISHNU} hybrid model calculations for the elliptic flow of 
$\pi$, p + $\bar{\rm p}$ and $\Lambda + \bar{\Lambda}$ (figure from~\cite{Abelev:2014pua}).
}
\label{fig:phi_ncq} 
\end{figure}
It is argued that the $\phi$-meson has a smaller hadronic interaction cross section, which would make it less affected by 
the late hadronic stage of the collision. It would therefore build up less radial flow than the other particles and this would
break the mass hierarchy of $v_2$.  In the left panel of Fig.~\ref{fig:phi_ncq} the {\sc VISHNU}~\cite{Song:2013qma} 
predictions for $v_2$ as 
function of transverse momentum are plotted for pions, protons and the $\phi$-meson. 
{\sc VISHNU} is a hybrid model calculation which matches a $(2+1)$-dimensional longitudinally boost-invariant viscous hydrodynamic calculation to a hadron cascade model ({\sc UrQMD}). 
In this hybrid model the Cooper-Frye algorithm is used to convert the hydrodynamical system into particle phase-space distributions on a switching surface of constant temperature $T_{\rm chem}$, 
assuming chemical equilibrium particle yields. 
The hadron cascade {\sc UrQMD} then propagates these particles until the kinetic freeze-out, $T_{\rm kin}$, 
where all interactions cease and all unstable resonances have decayed.
In such a hybrid model the difference in $v_2$ due to different hadronic interaction cross sections can be calculated. 
The {\sc VISHNU} calculations indeed clearly show that the smaller hadronic cross section leads to a 
larger $v_2$ below 2 GeV/$c$ for the $\phi$-meson compared to the proton $v_2$.
In the same figure the $\phi$-meson $v_2$ is also plotted for hydrodynamic calculations with two decoupling temperatures, 
$T_{\rm chem}$ and $T_{\rm kin}$ at 165 and 100 MeV, respectively.
This comparison shows that the $\phi$-meson $v_2$ indeed does not change significantly anymore in {\sc UrQMD}, 
which models the hadronic contribution. 
Unfortunately, the uncertainties in the ALICE $\phi$-meson $v_2$ are currently still too large at low-$\pt$ to 
constrain the hadronic contribution. 

However, not only for the $\phi$-meson $v_2$ there 
is a difference between a hybrid and full hydrodynamical model calculations. 
Strange baryons also have a somewhat smaller hadronic cross section, and even though for the $\Lambda$ this difference is 
much smaller it does break the mass hierarchy compared to the proton in the same {\sc VISHNU} calculation. 
These calculations compared to the ALICE proton and $\Lambda$ $v_2$ measurements are shown in the 
right panel of Fig.~\ref{fig:phi_ncq}.
The ALICE measurements of $\Lambda$ $v_2$ are much more precise than the $\phi$-meson $v_2$ and 
clearly show that the mass hierarchy is preserved~\cite{Abelev:2014pua}, 
in clear contrast with {\sc VISHNU} model predictions. 

At intermediate $\pt$ the $v_2$ values of baryons (i.e. protons, $\Lambda$s, $\Xi$s, $\Omega$s 
and their antiparticles) become very similar. The meson $v_2$, $\pi$ and $K$, are also the same within uncertainties.
The meson $v_2$ are, however, well below the baryons. 
This behaviour at intermediate $\pt$ is thus very different compared to the mass ordering at low-$\pt$. 
The transverse momentum where the baryon and the pion $v_2$ cross depends on centrality, moving to higher $\pt$ 
for more central collisions. 
At these intermediate $\pt$ it is argued that the baryons and mesons are formed via 
constituent quark coalescence~\cite{Greco:2003xt,Greco:2003mm}. 
If indeed the baryon and mesons are formed 
in such a simplified coalescence picture the $v_2$ of the particle would only depend on the number of its 
constituent quarks~\cite{Molnar:2003ff}. 
This implies that all baryons have the same $v_2$ (independent of their mass) and 
also all mesons have a single value of $v_2$. 
The ratio between baryon and mesons $v_2$ should be $3/2$,  the ratio of the number of constituent quarks. 
Figure~\ref{fig:identifiedv2} shows that $\phi$-meson breaks this scaling and is at intermediate $\pt$ for the more 
central collisions closer to the baryons while for the more peripheral closer to the mesons.

\begin{figure}[h!]
\includegraphics[width=0.5\textwidth]{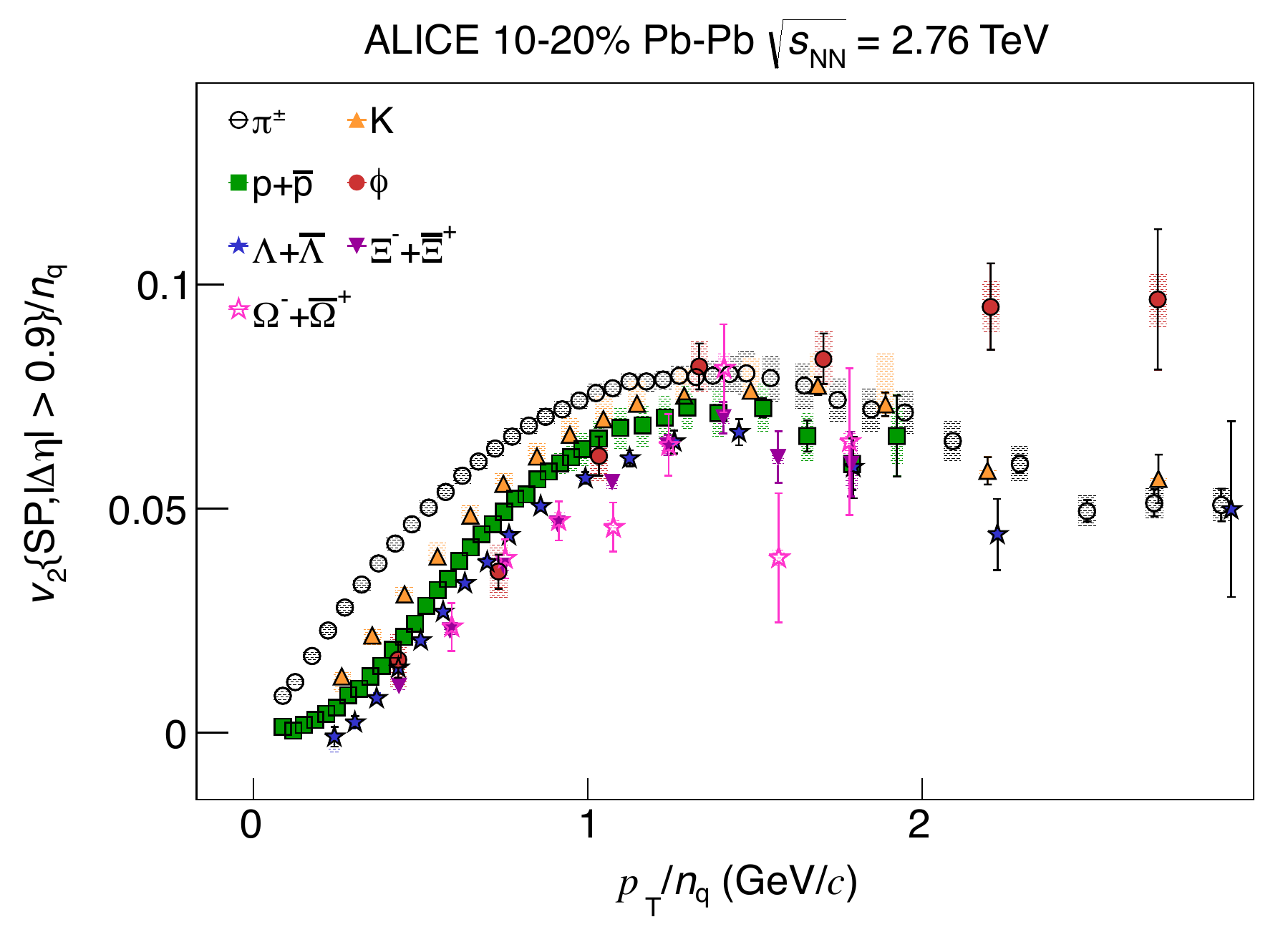}
\includegraphics[width=0.5\textwidth]{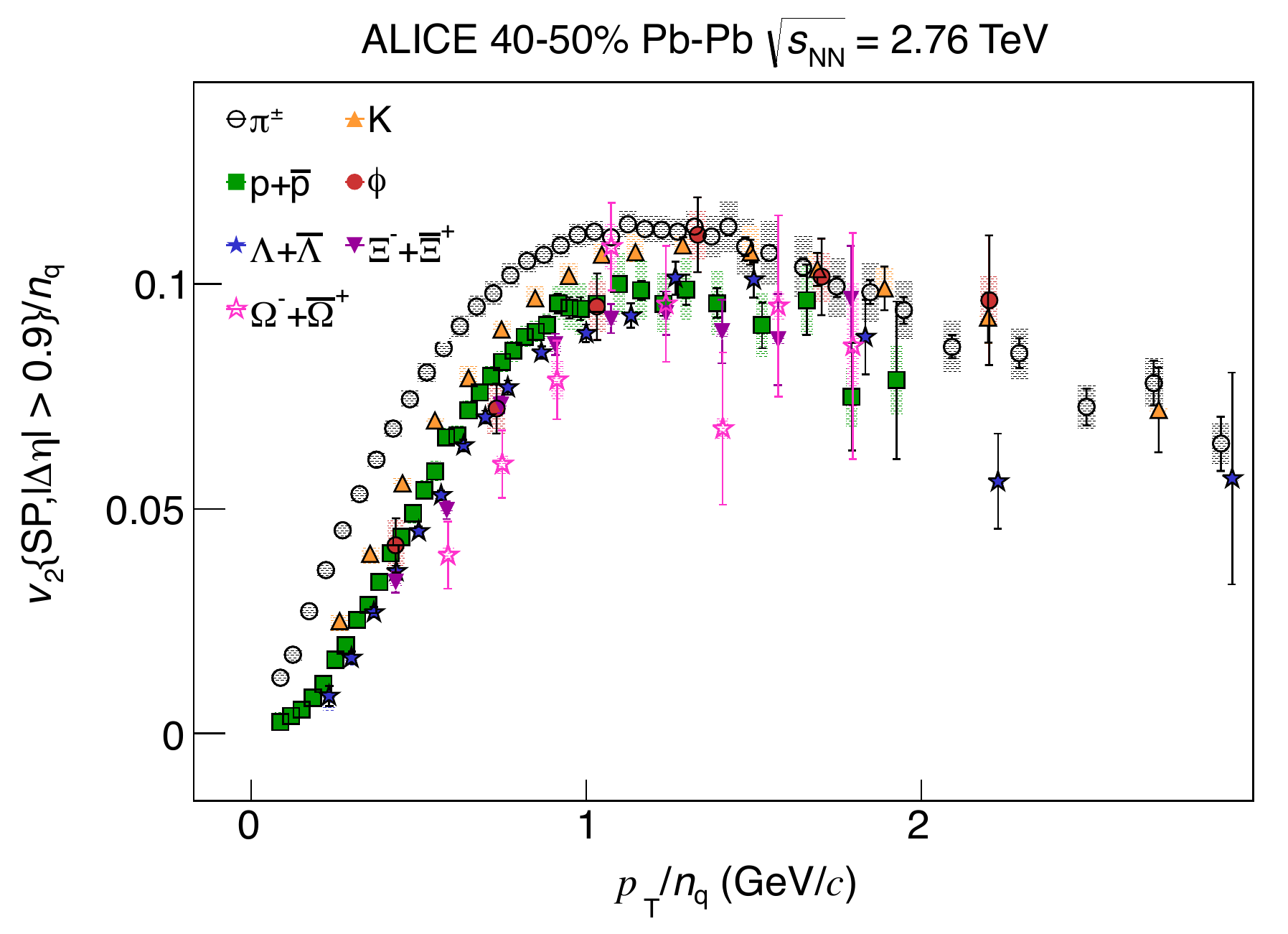}
\caption{
$\pt$-differential elliptic flow of identified particles scaled by the number of constituent quarks 
for the 10--20\% (left) and 40--50\% (right) centrality class
(figures from~\cite{Abelev:2014pua}).
}
\label{fig:v2_ncq} 
\end{figure}
To test the number of constituent quark scaling for the other particles, $v_2/n_q$ 
is plotted versus $\pt/n_q$ in Fig.~\ref{fig:v2_ncq}, where $n_q$ is the number of constituent quarks.
The left and right panels show the scaling for 10--20\% and 40--50\% centrality, respectively. 
The data shows that number of constituent quark scaling is only approximate for the baryons and mesons.

\subsection{Anisotropic flow fluctuations}
\label{fluctuations}

Due to fluctuations in the initial density profile the initial spatial geometry does not have a smooth almond shape but, 
instead, a more complex spatial geometry which may possess also odd harmonic symmetry planes. 
These fluctuations contribute to the measurements of elliptic flow and in addition, are predicted to give rise 
to odd harmonics like triangular flow $v_3$. 

The difference between $v_{n}\{2\}$ and $v_{n}\{4\}$ is sensitive to nonflow and the event-by-event $v_n$ fluctuations.
From Eq.~\ref{fluctuations1} and the data in the right panel of Fig.~\ref{fig:integratedv2}, 
under the assumption that the nonflow contribution $\delta_{2,2}$ is understood and $\sigma_{v_2} \ll \bar{v}_2$,  
$\sigma_{v_2}$ can be obtained.  
\begin{figure}[h!]
\includegraphics[width=0.5\textwidth]{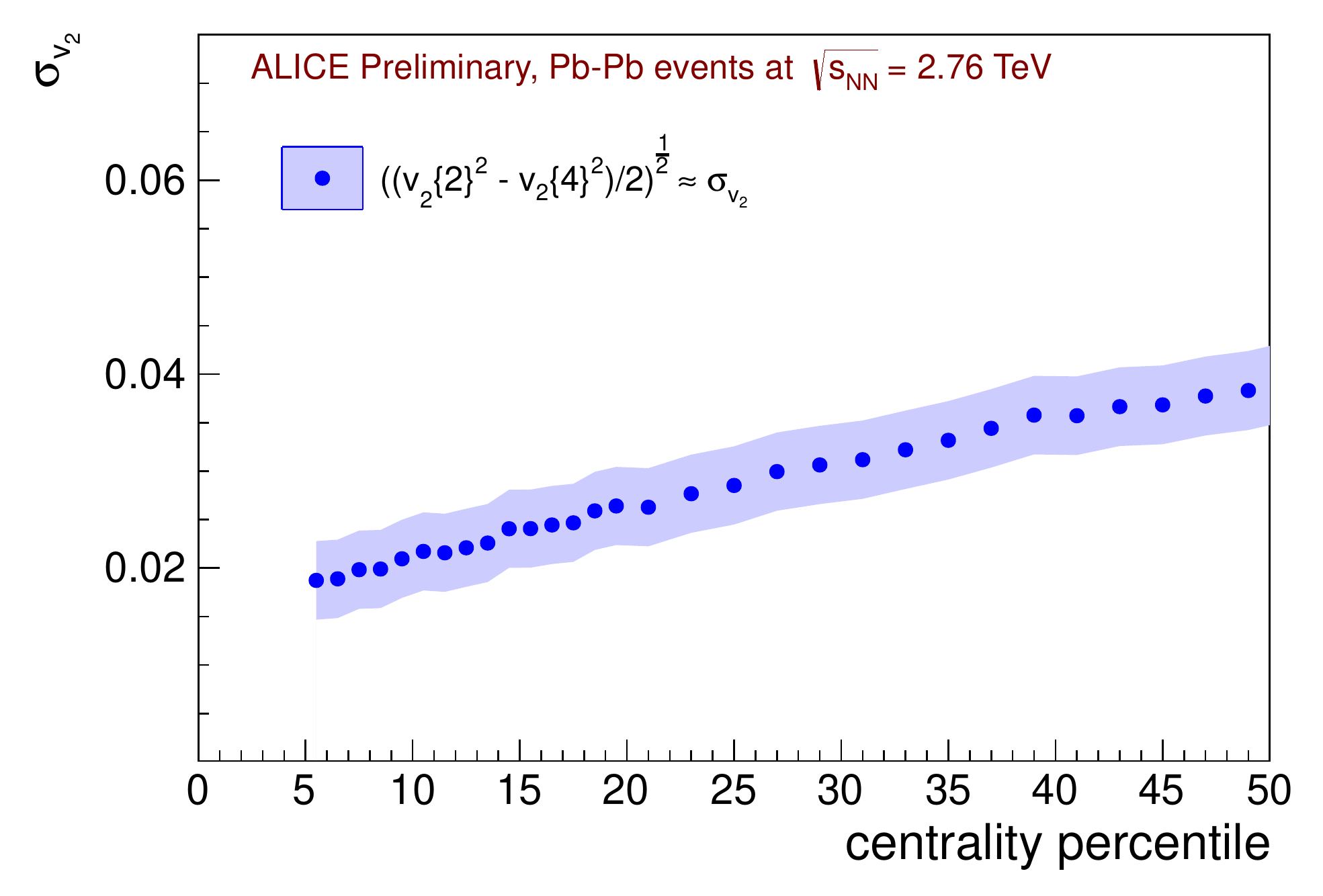}
\includegraphics[width=0.5\textwidth]{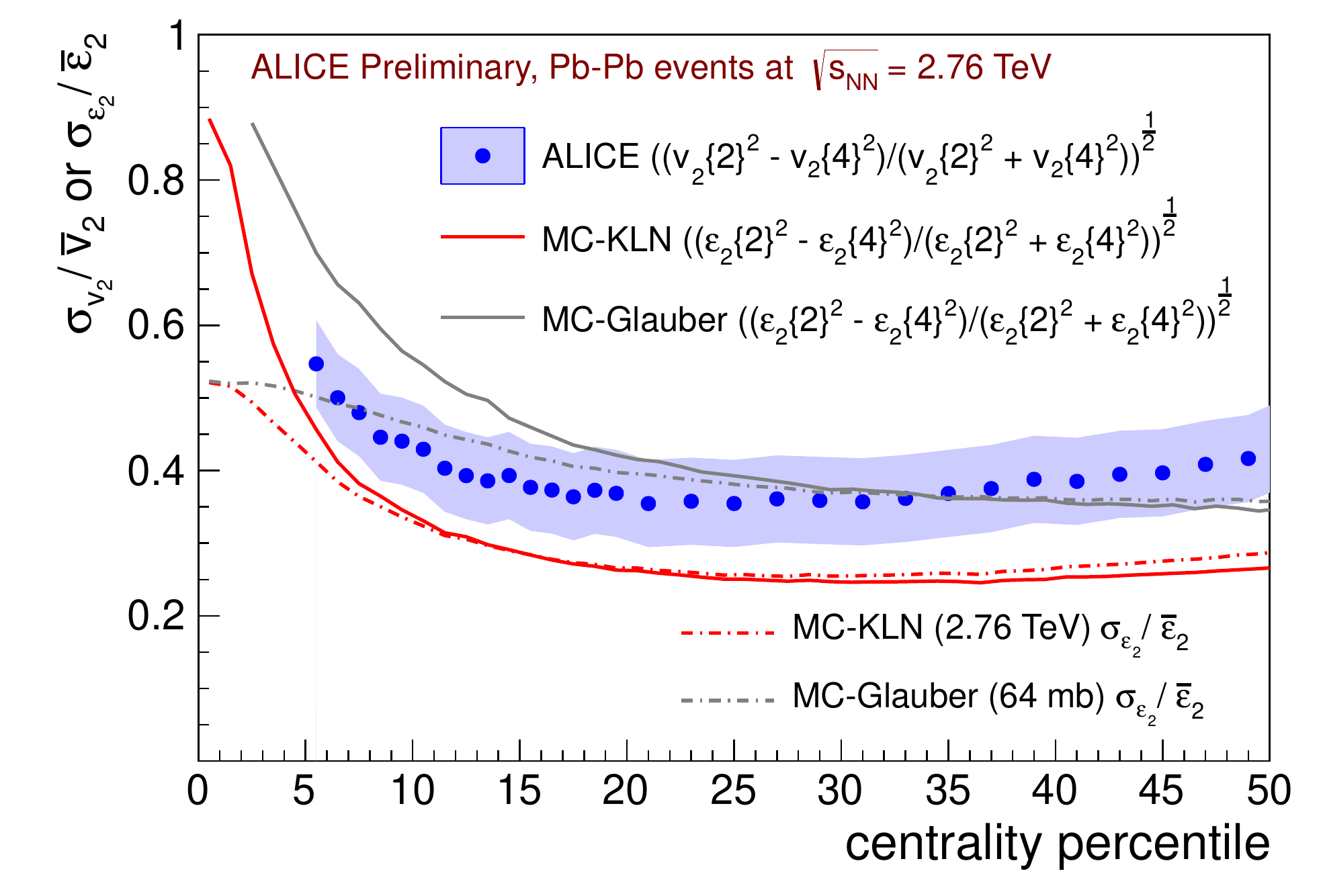}
\caption{
Left: The magnitude of the event-by-event elliptic flow fluctuations versus collision centrality.
Right: Relative event-by-event elliptic flow fluctuations versus collision centrality 
(figures from~\cite{Collaboration:2011yba}).
}
\label{fig:sigmav2} 
\end{figure}
The results are plotted in Fig.~\ref{fig:sigmav2} (left), together with the ratio $\sigma_{v_2} / \bar{v}_2$ (right). 
The ratio $\sigma_{v_2} / \bar{v}_2$ is large $\sim 40$\%, similar to measurements at RHIC~\cite{Alver:2007qw}. 
For these large fluctuations $\sigma_{v_2} \ll \bar{v}_2$ is not true, which implies 
that Eq.~\ref{fluctuations1} is not generally applicable.

The magnitude of the elliptic flow is proportional to $\varepsilon_2$ of the initial density profile, and 
hydrodynamical model calculations have also shown that the event-by-event fluctuations in 
the elliptic flow are proportional to that in $\varepsilon_2$. 
The right panel of Fig.~\ref{fig:sigmav2} shows therefore the measured 
ratio $((v_2\{2\}^2 - v_2\{4\}^2)/(v_2\{2\}^2 + v_2\{4\}^2))^{1/2}$ compared to 
$((\varepsilon_2\{2\}^2 - \varepsilon_2\{4\}^2)/(\varepsilon_2\{2\}^2 + \varepsilon_2\{4\}^2))^{1/2}$ from a 
MC-Glauber and MC-KLN model. 
This comparison does not depend on the assumption that $\sigma_{v_2} \ll \bar{v}_2$ or 
$\sigma_{\varepsilon_2} \ll \bar{\varepsilon}_2$.
The MC-KLN under-predicts the data whereas the MC-Glauber over-predicts the data for more central collisions. 
To investigate to which extent the ratio plotted in Fig~\ref{fig:sigmav2} (right) 
represents $\sigma_{\varepsilon_2} / \bar{\varepsilon}_2$ 
this ratio is calculated directly from the distributions generated by the two models, that is, 
without using Eq.~\ref{fluctuations1}. For mid-central collisions $\sigma_{\varepsilon_2} / \bar{\varepsilon}_2$ and 
$((\varepsilon_2\{2\}^2 - \varepsilon_2\{4\}^2)/(\varepsilon_2\{2\}^2 + \varepsilon_2\{4\}^2))^{1/2}$ are very similar in
both MC models of the initial state.

Recently it was realised that the odd harmonics are particularly sensitive to both $\eta/s$ and the initial conditions, 
which generated strong theoretical and experimental 
interest~\cite{Takahashi:2009na,Sorensen:2010zq, Alver:2010gr,Alver:2010dn,Floerchinger:2013vua,
Floerchinger:2014fta,ALICE:2011ab,Adare:2011tg}. 
The odd harmonics also give rise to long range 
correlations in the longitudinal direction. These correlations were first observed at RHIC and initially 
interpreted as due to jet modifications in the hot and dense matter (i.e. Mach Cones).

\begin{figure}[h!]
\includegraphics[width=0.5\textwidth]{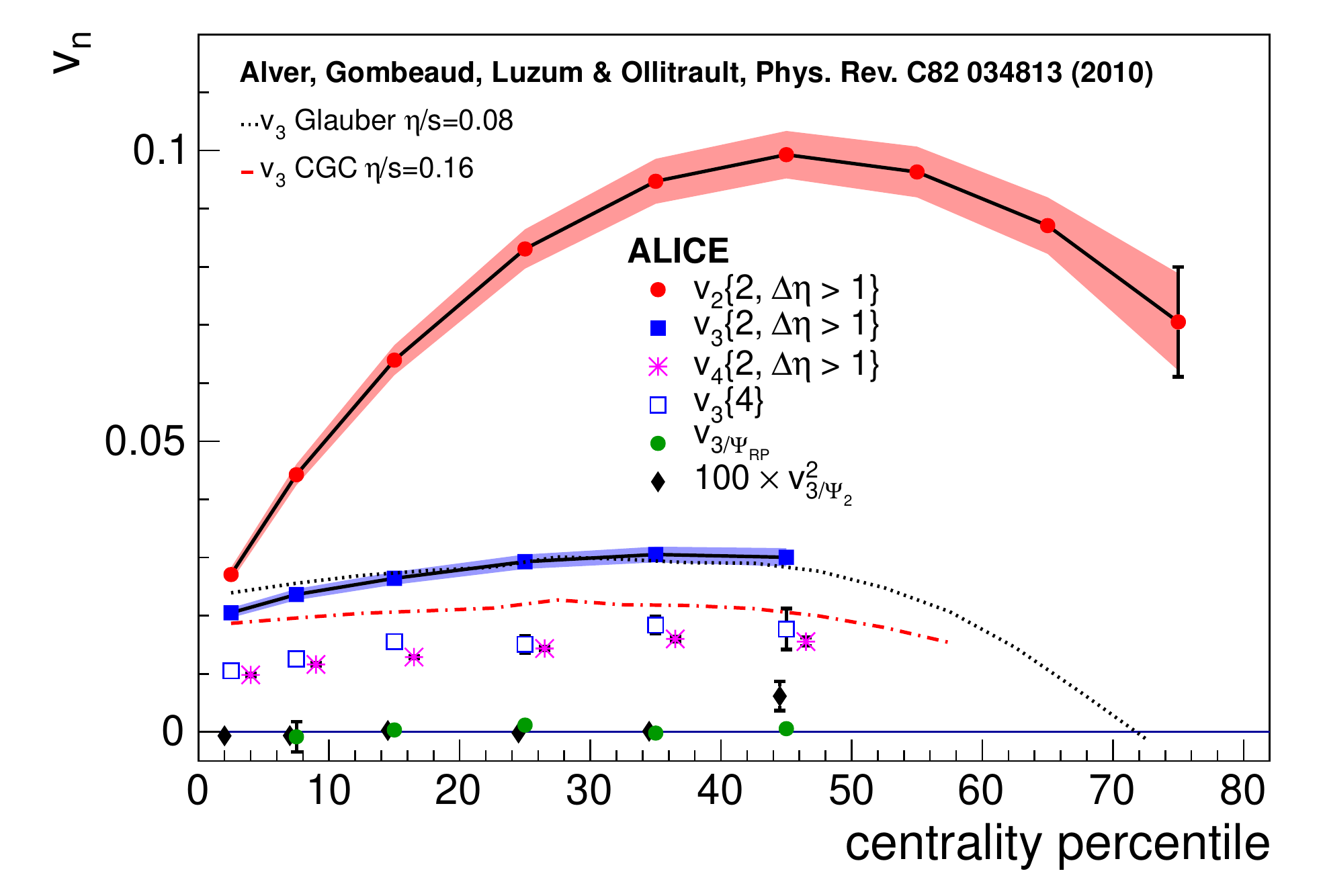}
\includegraphics[width=0.5\textwidth]{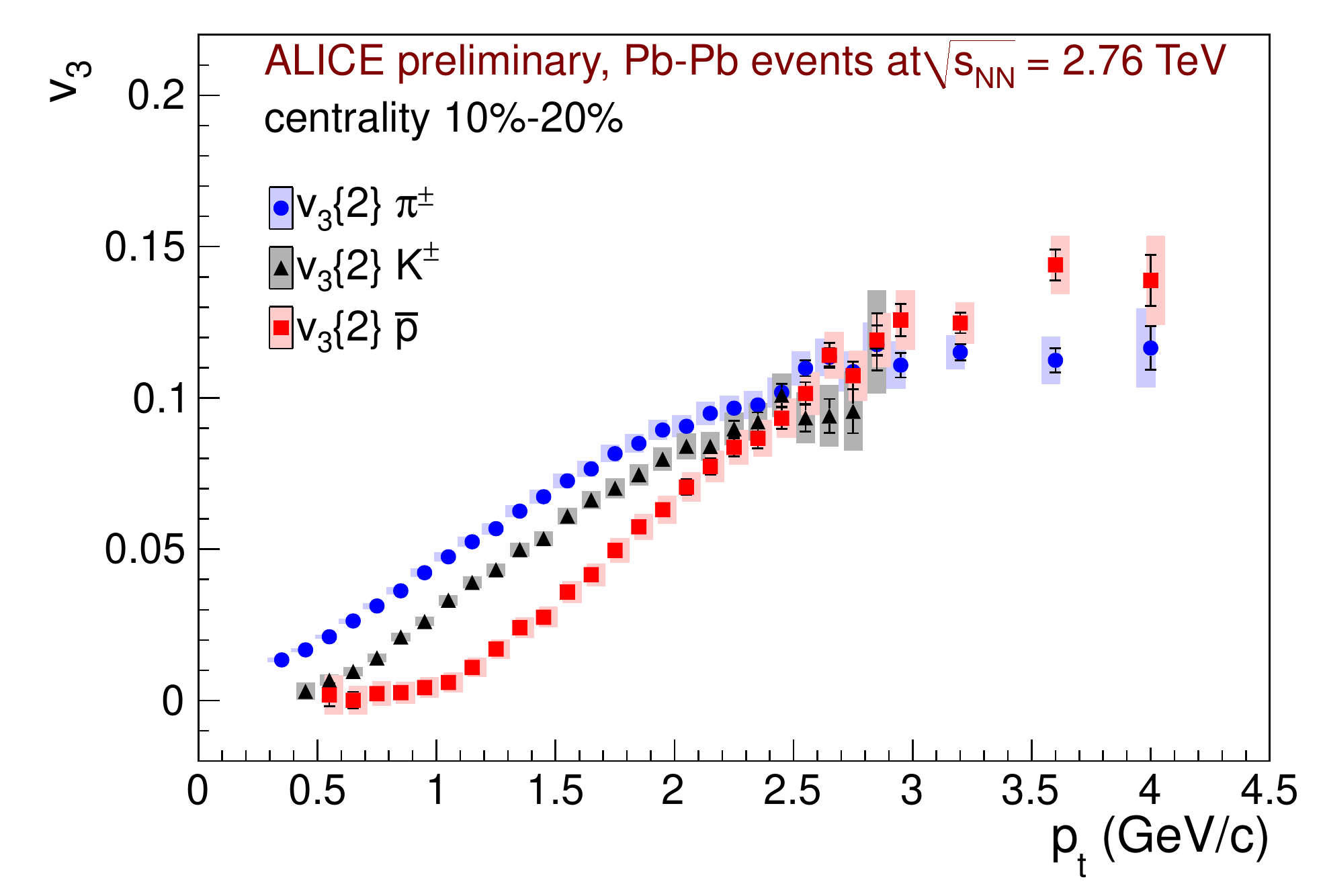}
\caption{
Left: Integrated $v_2$, $v_3$ and $v_4$, full and open squares show
$v_{3}\{2\}$ and $v_{3}\{4\}$ respectively. In addition we show $v_{3/\Psi_{2}}^{2}$ and $v_{3/\Psi_{\rm RP}}$, 
which represent the triangular flow measured relative to the second order event plane and the reaction plane, respectively.
The dashed curves are hydrodynamical predictions~\cite{Alver:2010dn} (Figure adapted from~\cite{ALICE:2011ab}) 
Right: The $\pt$-differential triangular flow for pions, kaons and antiprotons 
(Figure taken from~\cite{Krzewicki:2011ee}).
}
\label{fig:triangularflow} 
\end{figure}
Figure~\ref{fig:triangularflow} (left) shows $v_2$, $v_3$ and $v_4$ integrated over $\pt$ 
as a function of centrality.  The $v_n\{2\}$ are measured correlating particles which are 
separated by at least one unit of pseudorapidity, and in addition corrected for the estimated remaining 
nonflow contribution based on model calculations. 
The total systematic uncertainty is shown as a band and fully includes this residual correction.
The measured $v_3$ is smaller than $v_2$ and does not depend strongly on centrality. 
The  hydrodynamic model calculation plotted in the figure show that $v_3$ is compatible with predictions 
for Pb-Pb collisions from a calculation with Glauber initial 
conditions and $\eta/s =0.08$ and larger than for MC-KLN CGC initial conditions with 
$\eta/s=0.16$~\cite{Alver:2010dn}, suggesting a small value of  
$\eta/s$ for the matter produced in these collisions. 
The $v_3$\{4\} is about a factor two smaller than the two-particle measurement 
which can not  be understood if the underlying fluctuations are distributed as a 2D 
Gaussian.  
Recent studies suggests an elliptic power distribution as an alternative underlying p.d.f., which
predicts a non-zero $v_{3}\{4\}$ and is compatible with the data~\cite{Yan:2013laa,Yan:2014afa}. 

For these event-by-event fluctuations of the spatial geometry, the symmetry plane $\Psi_3$ is expected to be  
almost uncorrelated with the reaction plane $\Psi_{\rm RP}$~\cite{Poskanzer:1998yz}, 
and with the elliptic flow plane $\Psi_2$.  
The correlations between
$\Psi_3$ and $\Psi_{\rm RP}$ is calculated using $v_{3/\Psi_{\rm RP}}=\left< \cos (3\varphi_1
-3\Psi_{\rm RP})\right>$ and the correlation between $\Psi_3$ and
$\Psi_{2}$ with a five-particle correlator $\left< \cos (3\varphi_1 +
3\varphi_2 - 2\varphi_3 - 2\varphi_4 - 2\varphi_5)\right>/v_{2}^{3}=v_{3/\Psi_2}^{2}$.
In the left panel of Fig.~\ref{fig:triangularflow} $v_{3/\Psi_{\rm RP}}$ and 
$v_{3/\Psi_{2}}^2$ are shown as a function of centrality. These correlations are
indeed, within uncertainties, consistent with zero as expected. 

The event-by-event $v_n$-distributions~\cite{Aad:2013xma} and 
the correlations between other planes also have been measured~\cite{Aad:2014fla} 
in ATLAS and they are described in detail in another contribution to this volume~\cite{Jia:2014jca}. 

In the right panel of Fig.~\ref{fig:triangularflow}, to investigate further the hydrodynamic origin of $v_3$, 
the $\pt$-differential $v_3$ of pions, kaons and antiprotons is plotted. 
It is seen that a similar mass splitting pattern as observed in elliptic flow is also clearly present in 
triangular flow as well, as is expected if both have the same hydrodynamic origin. 

\begin{figure}[h!]
\includegraphics[width=0.52\textwidth]{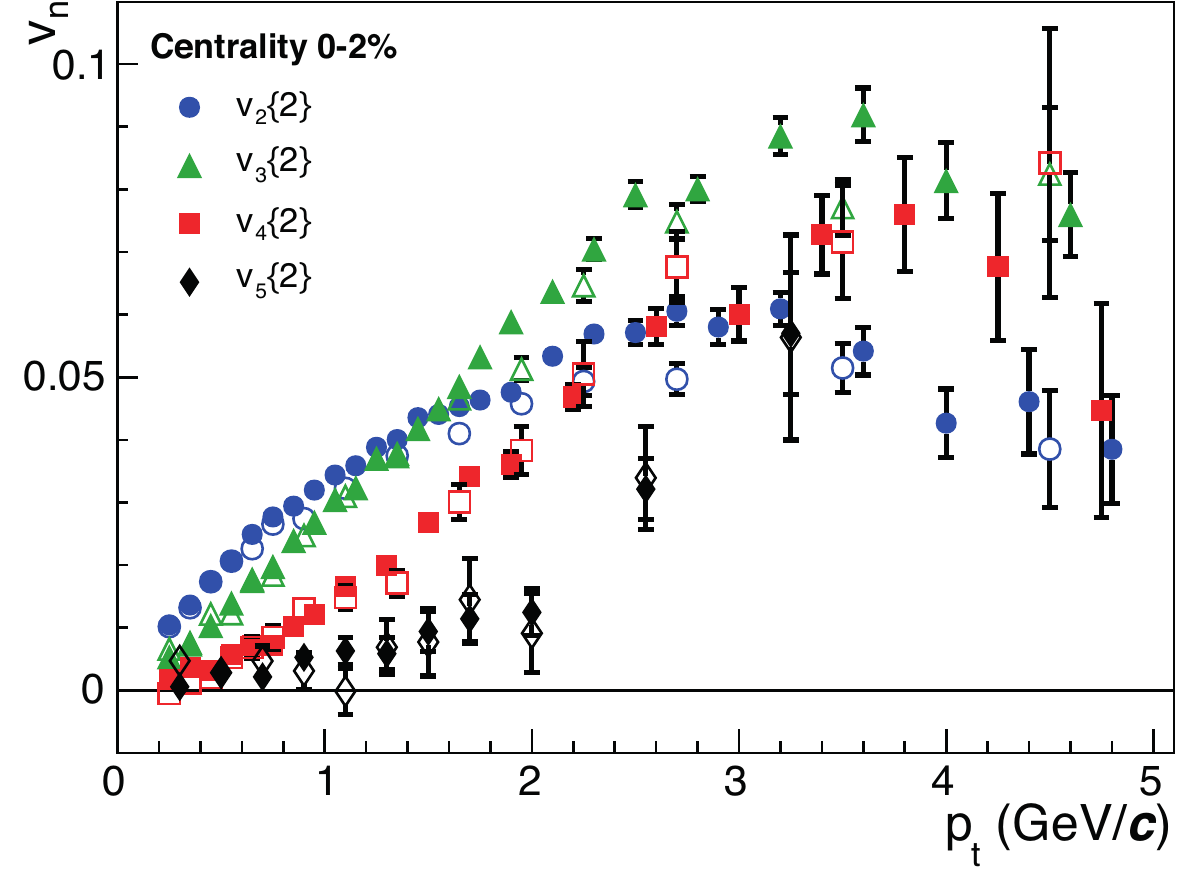}
\includegraphics[width=0.48\textwidth]{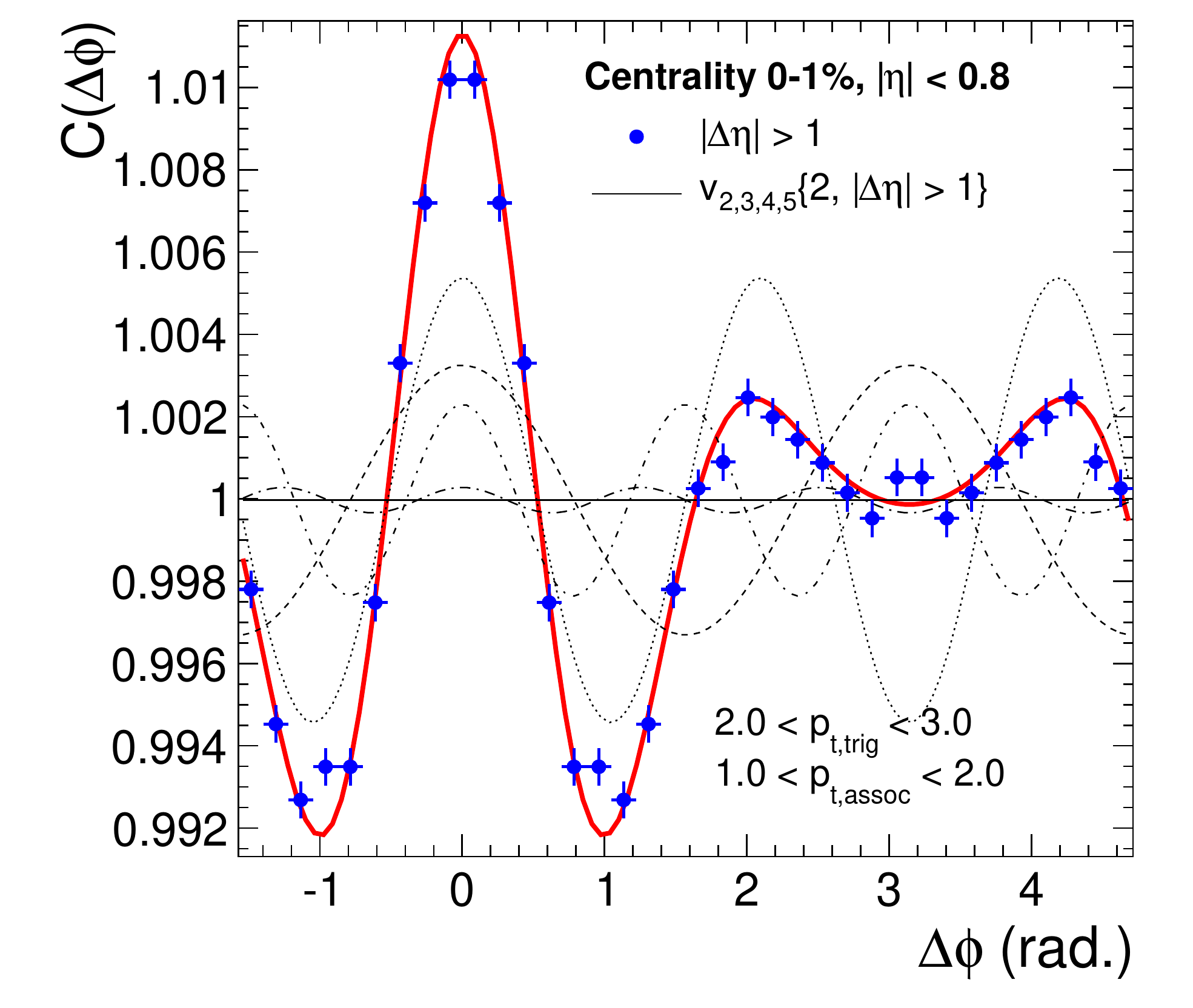}
\caption{
Left: $v_2$, $v_3$, $v_4$, $v_5$ as a function of
    transverse momentum. The full, open symbols 
    are for  $\Delta\eta > 0.2$ and  $\Delta\eta > 1.0$, respectively. 
Right: The two-particle azimuthal correlation, 
    measured in $0 < \Delta\phi < \pi$ and shown symmetrized over $2\pi$, between a 
    trigger particle with $2 < \pt < 3$ GeV/$c$ and an
    associated particle with $1 < \pt < 2$ GeV/$c$ for the 0--1\%
    centrality class. The solid red line shows the sum of the measured 
    anisotropic flow Fourier coefficients $v_2$, $v_3$, $v_4$ and $v_5$ (dashed lines). 
    Figures taken from~\cite{ALICE:2011ab}.
}
\label{fig:machcones} 
\end{figure}
The integrated elliptic flow is dominant compared to the other $v_n$ for all centralities due to the almond 
shape of the initial spatial density profile. For the most central collisions all the $v_n$ approach each other in 
magnitude, as is expected when the fluctuations start to dominate all $\varepsilon_n$. For the most 0--2\% central 
collisions the  $\pt$-differential $v_n$ is plotted in the left panel of Fig.~\ref{fig:machcones}. 
For these very central collisions $v_3$ becomes equal to $v_2$ already below 2 GeV/$c$, 
while above 3 GeV/$c$ $v_3$, $v_4$ and $v_5$ are all at least equal or larger than $v_2$ (this behaviour is also 
confirmed by CMS measurements~\cite{CMS:2013bza}).

For these very central collisions the large contribution of $v_3$ is responsible for an interesting structure in the two-particle azimuthal correlation 
between so called triggered and associated particles. The two-particle azimuthal correlations are measured by
calculating: 
\begin{equation}
C(\Delta\phi) \equiv \frac{N_{\rm mixed}}{N_{\rm same}} \frac{{\rm d}N_{\rm same}/{\rm d}\Delta\phi}{{\rm d}N_{\rm mixed}/{\rm d}\Delta\phi} ,
\end{equation}
where $\Delta\phi =\phi_{trig} -\phi_{assoc}$, d$N_{\rm same}$/d$\Delta\phi$ (d$N_{\rm mixed}$/d$\Delta\phi$) is the 
number of associated particles as function of $\Delta\phi$ within the same (different) event, 
and $N_{\rm same}$ ($N_{\rm mixed}$) the total number of associated particles in d$N_{\rm same}$/d$\Delta\phi$ (d$N_{\rm mixed}$/d$\Delta\phi$).

Figure~\ref{fig:machcones} (right) shows the azimuthal correlation observed in very central collisions 
0--1\%, for trigger particles in
the range $2 < \pt < 3$ GeV/$c$ with associated particles in $1 < \pt < 2$ GeV/$c$ for pairs in $|\Delta\eta| > 1$.  
A clear doubly-peaked correlation structure centered
opposite to the trigger particle is observed.  This structure also was observed at
lower energies in broader centrality bins~\cite{Adare:2008ae,Aggarwal:2010rf}, but only after
subtraction of the elliptic flow component.
In the past, this two-peak structure was interpreted as an indication for
various jet-medium modifications~\cite{Adare:2008ae,Aggarwal:2010rf} instead of 
a manifestation of triangular flow~\cite{Alver:2010gr,Alver:2010dn,Teaney:2010vd,Luzum:2010sp}.

\begin{figure}[htb]
 \includegraphics[width=0.5\textwidth]{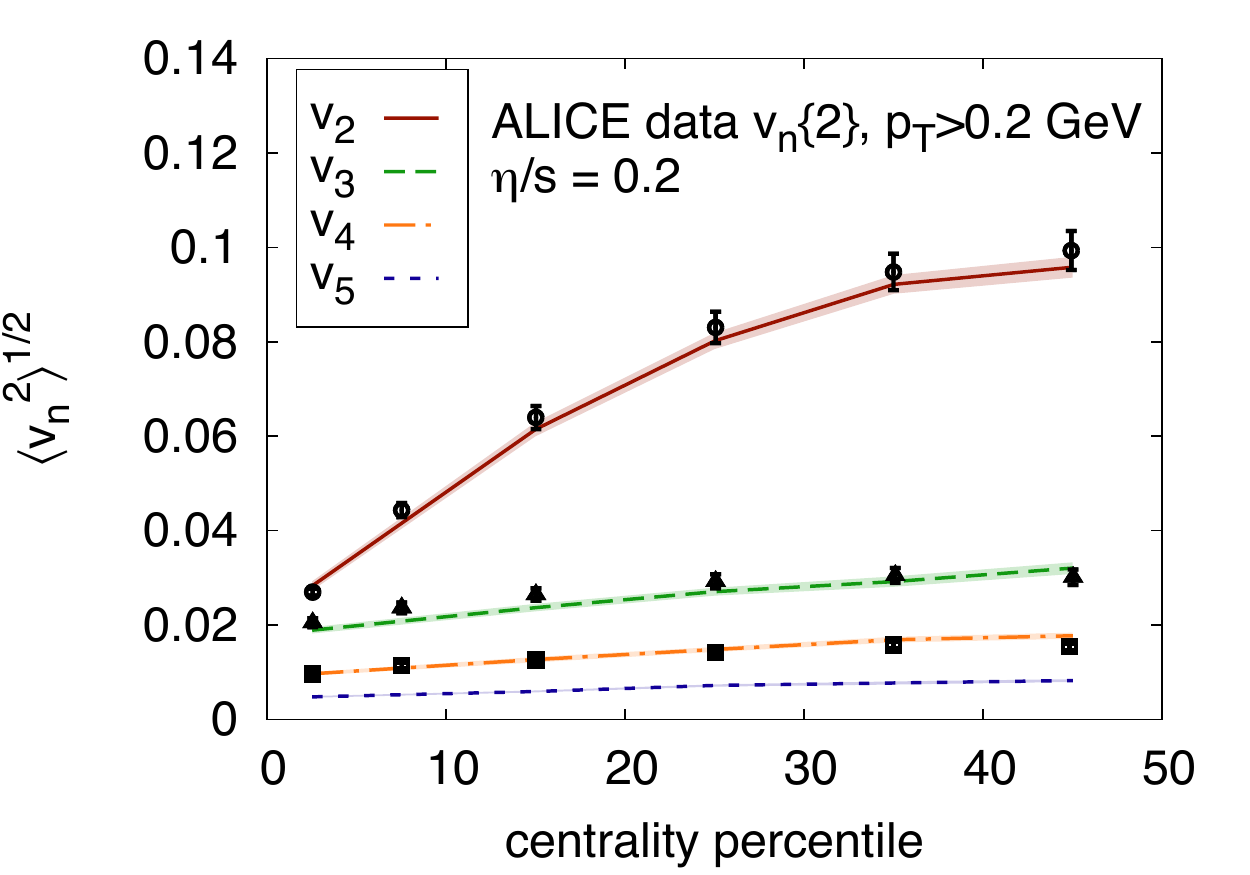}
 \includegraphics[width=0.5\textwidth]{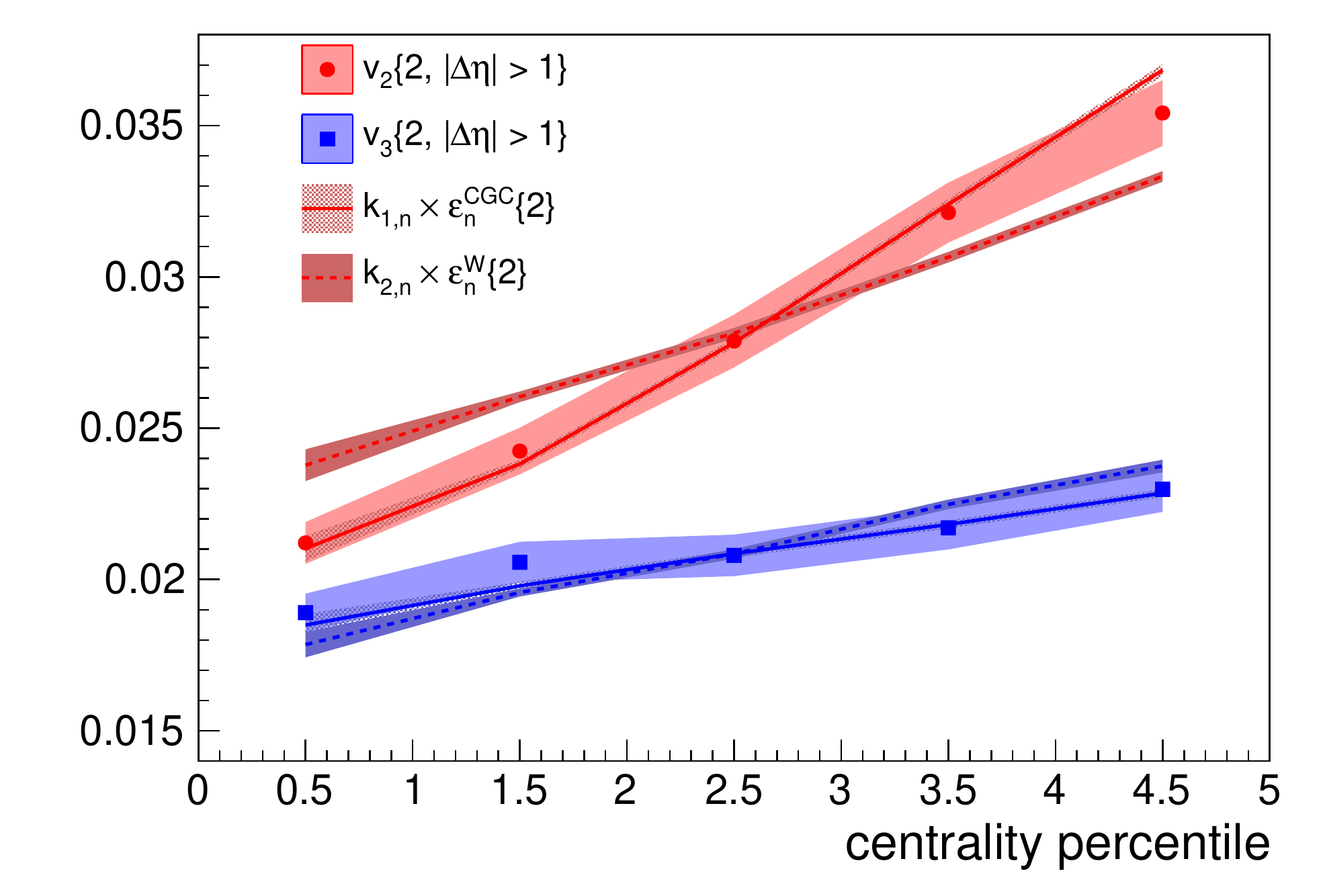}
 \caption{
      Left: The centrality dependence of $v_n\{2\}$ from 2.76 A TeV Pb+Pb collisions measured by 
      ALICE~\cite{ALICE:2011ab} 
      compared to viscous hydrodynamic model calculations~\cite{Gale:2012rq}  (Figure from~\cite{Gale:2012rq}).
      Right: $v_2$ and $v_3$ as a function of centrality for the 5\% most central collisions compared to calculations
      of the spatial eccentricities, $\varepsilon_{n}^W$\{2\} and $\varepsilon_{n}^{\rm CGC}$\{2\}.
      The eccentricities have been scaled to match the 2-3\% data using $k_1$ and $k_2$ 
      (Figure from~\cite{ALICE:2011ab}).
  \label{fig:verycentral} 
  }
\end{figure}
The hydrodynamical description of the measured $v_n$ is already rather well understood. 
Figure~\ref{fig:verycentral} shows a comparison 
between the measured $v_2$, $v_3$ and $v_4$ by ALICE~\cite{ALICE:2011ab} and a hydrodynamical model 
calculation using IP-Glasma initial conditions, 
together with a kinematic viscosity $\eta/s = 0.2$. These calculations provide a good description of the 
presently available data for $v_n$ and the $\pt$-differential $v_n$ (not shown here).

The $v_n$ depend on both the details of the spatial density distribution and on the kinetic viscosity. 
Detailed measurements of all the $v_n$ versus 
centrality provide a way to disentangle both contributions. 
There is in particular good sensitivity in the most central collisions, because 
viscous correction do not change much and therefore one is more 
sensitive to the initial spatial density distributions and its fluctuations. 
In Fig.~\ref{fig:verycentral} (right) $v_2$\{2\} and $v_3$\{2\} are
plotted in 1\% centrality bins for the 5\% most central collisions. 
The $v_3$\{2\} do not change much versus centrality (as
would be expected if $v_3$ is dominated by event-by-event fluctuations
of the initial geometry) while, the $v_2$\{2\}
increase by about 60\%.  Compared to the centrality dependence of the eccentricities $\varepsilon_n$\{2\}
for initial conditions from MC-KLN CGC and MC-Glauber, it is seen   
that the weak dependence of $v_3$\{2\} is described by both models of the initial stage 
while the relative strong dependence of $v_2$\{2\} on centrality is only described for the 
MC-KLN CGC initial conditions.

\subsection{Event-shape engineering}
\label{engineering}
Controlling the initial conditions in heavy-ion collisions is of the utmost importance to learn about 
the properties of high density hot QCD matter.  To experimentally control part of the overlap geometry we 
currently categorise the collisions by their collision centrality or use different collision systems. 
However with better understanding of the role of the fluctuations in the initial energy density 
distribution a new tool has become available to control the initial collision geometry. 

\begin{figure}[htb]
 \includegraphics[width=0.5\textwidth]{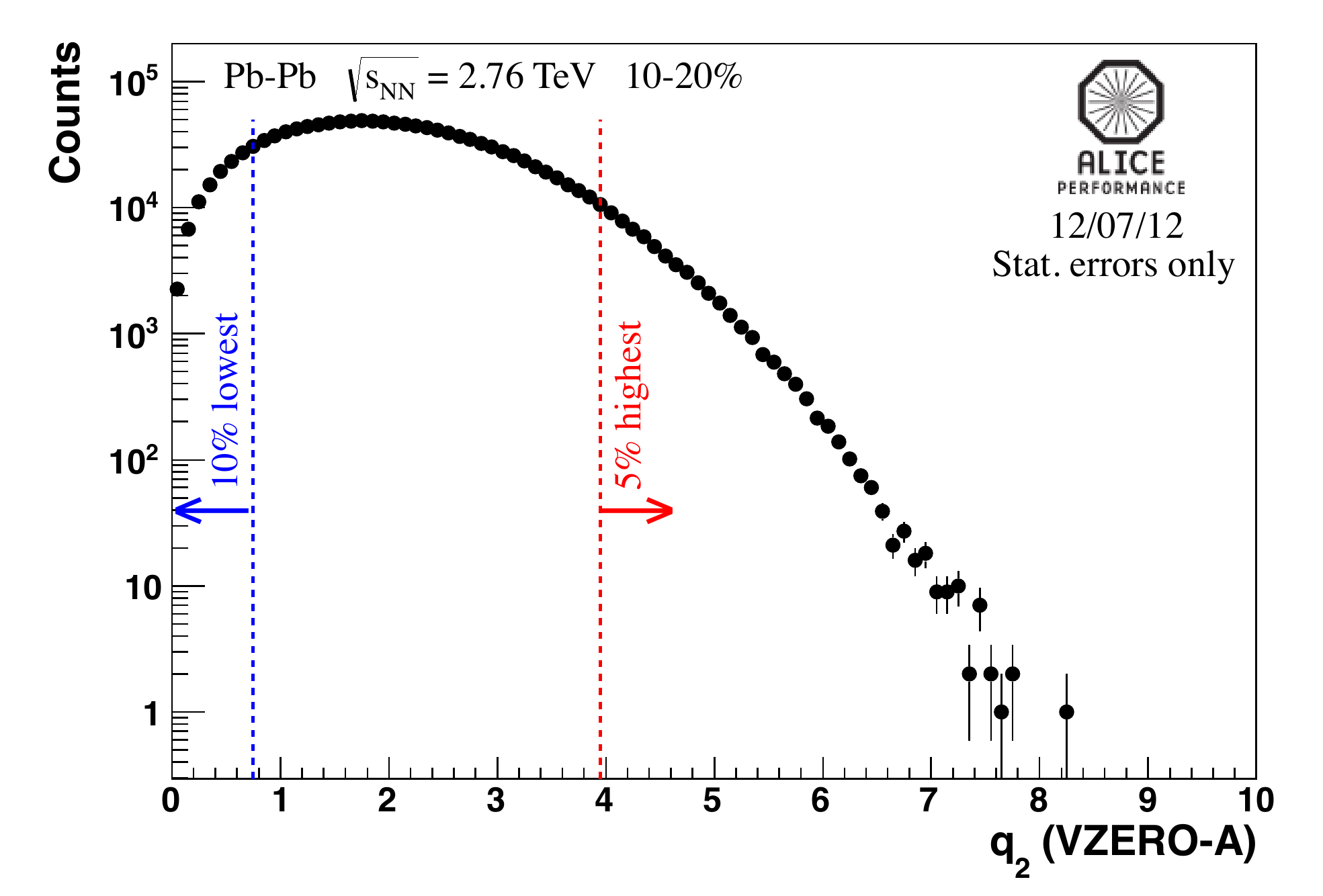}
 \includegraphics[width=0.5\textwidth]{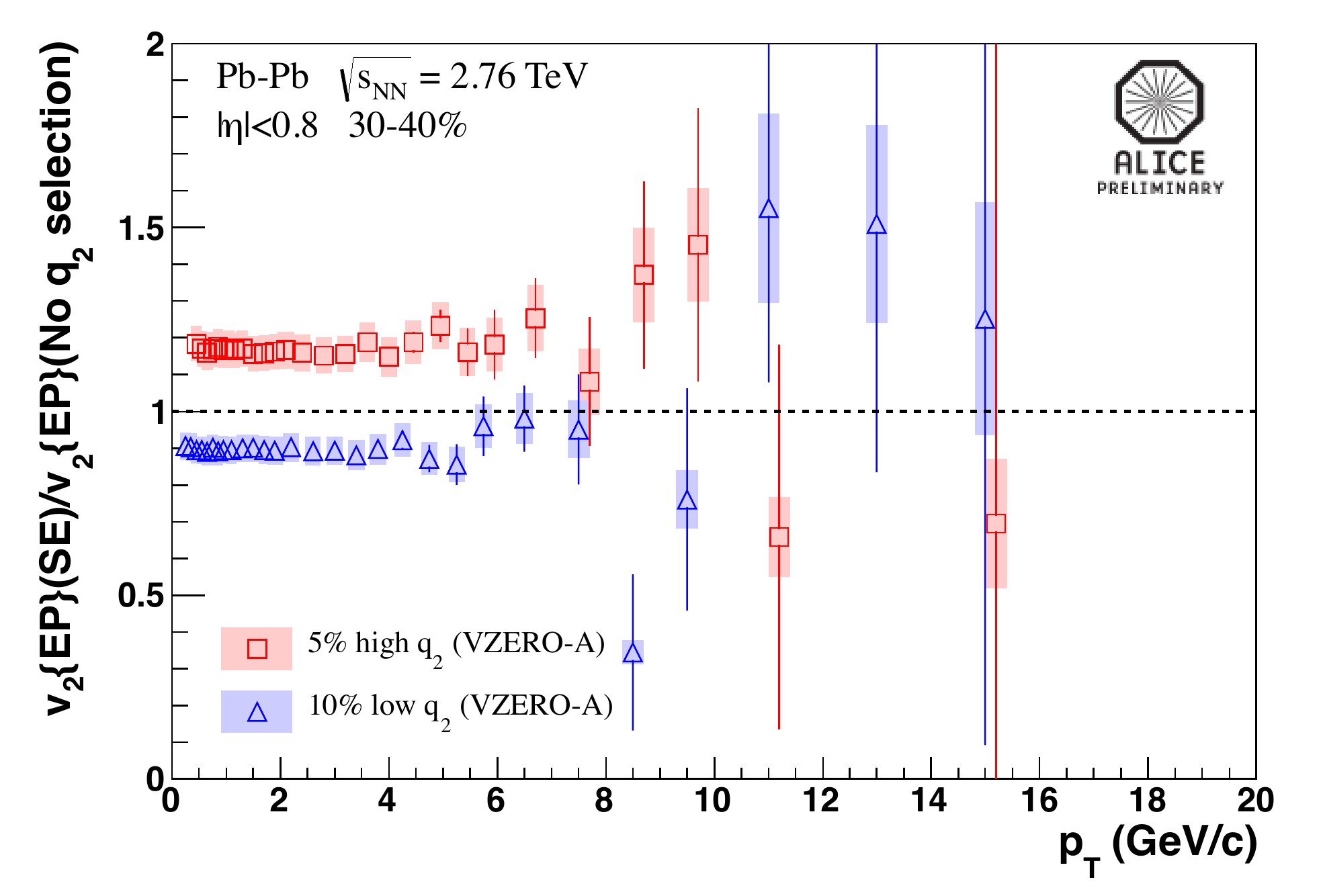}
 \caption{
      Left: Distribution of $q_2$ from the VZERO-A ($2.8 \le \eta \le 5.1$) for the 10--20\% centrality 
      (Figure from~\cite{Dobrin:2012zx}).
      Right: The ratio of unidentified charged particle $v_2$ between event 
     shape selected and unbiased events (Figure from~\cite{Dobrin:2012zx}).
  \label{fig:eventshape} 
  }
\end{figure}
At fixed collision centrality, due to fluctuations in the initial spatial distribution, the $v_n$ differ event-by-event. 
This can be used to select events with significantly larger or smaller values of $v_n$ compared to the 
average~\cite{Schukraft:2012ah}. 
In the left panel of Fig.~\ref{fig:eventshape} the so-called reduced elliptic flow vector $q_2$ distribution is plotted. 
The reduced flow vector is defined as:
\begin{equation}
q_n = Q_n / \sqrt{M},
\label{reducedflowvector}
\end{equation}
where $Q_n$ is the flow vector defined in Eq.~\ref{Qvector} and $M$ is the multiplicity of the particles used for the flow vector.
The nonflow contributions to the measurements are suppressed by choosing the reduced flow vector with a large (pseudo) 
rapidity separation with respect to where the anisotropic flow or other observables are measured. 
From the distribution of the reduced elliptic flow vector we can now select the events with the 5\% highest and 10\% 
lowest $q_2$ which have a different event shape. 
This event-shape engineering of the data sample offers new possibilities to test 
the properties of the hot and dense QCD matter~\cite{Schukraft:2012ah,Dobrin:2012zx}. 

The right-panel of Fig.~\ref{fig:eventshape} shows the charged particle $\pt$-differential $v_2$ 
ratio of event shape selected and unbiased events.
As expected for the events with the 
highest and lowest $q_2$ the measured $v_2(\pt)$ is also higher and lower, respectively. That the ratio of $v_2(\pt)$ is flat indicates that 
the flow fluctuations have a common origin at least up to 6 GeV/$c$. 

In addition to measurements of $v_n$, this event-shape engineering can be used to understand better many other observables.
Examples are azimuthally sensitive femtoscopy and also observables such as final state $\pt$ fluctuations, which both clearly would 
profit from selecting events with extreme values of the anisotropy.
However also for observables where $v_n$ is one of the main background contributions, event-shape engineering is 
an important tool and an example of this is the proposed chiral magnetic effect.
Because the chiral magnetic effect should hardly depend on the fluctuations in the initial spatial density distribution the background $v_n$ contribution 
could be estimated from measurements in events with the same collision centrality but different event-shape selections.
First applications of this new technique for various observables were recently 
presented~\cite{Dobrin:2012zx,Lacey:2013eia,Huo:2013qma,QM2014}.  

\subsection{Anisotropic flow in proton-nucleus collisions?}
\label{small_systems}

Azimuthal correlations in proton-proton and proton-nucleus collisions are dominated by near- and away-side jet peaks.
In nucleus-nucleus collisions additional correlations due to anisotropic flow are clearly visible. These anisotropic flow 
correlations quickly become dominant when we look at correlations between particles with a large separation in 
pseudo-rapidity.
However the observation of similar but smaller long range $\Delta \eta$ azimuthal 
correlations for very high multiplicity proton-proton collision came as a surprise~\cite{Khachatryan:2010gv}. 
Because these very high multiplicity events are extremely rare it is 
currently not clear if this long range $\Delta \eta$ structure, so called ridge, seen in proton-proton and nucleus-nucleus 
collisions has a common origin.

Data from a dedicated proton-nucleus (p-Pb) run enabled the LHC experiments to collect enough data to investigate  
if such a ridge also is visible in p-Pb and after that, to investigate in more detail what the origins of these correlations are.

In heavy-ion collisions, events are categorized according to their centrality based on the number of produced particles. 
In p-Pb the events are also subdivided into different event classes according to the number of produced particles, 
however for these collisions it is not clear how well this selection corresponds to the centrality of the collision. 
In ALICE the event classes are defined based on the multiplicity measured at forward pseudo-rapidity and denoted 
60--100\%, 40--60\%, 20--40\% and 0--20\% from the lowest to the highest multiplicity.

\begin{figure}[htb]
 \includegraphics[width=0.5\textwidth]{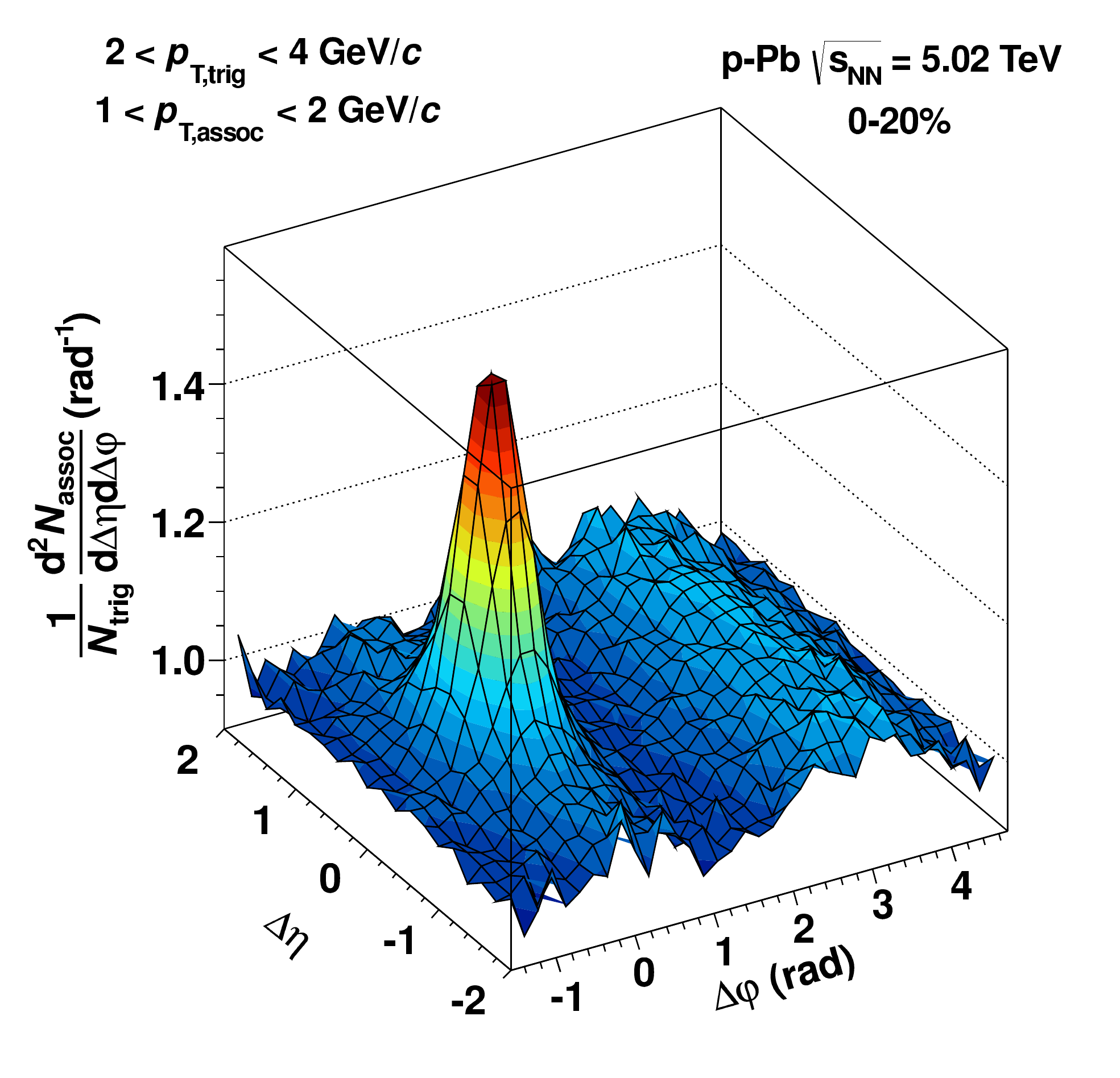}
 \includegraphics[width=0.5\textwidth]{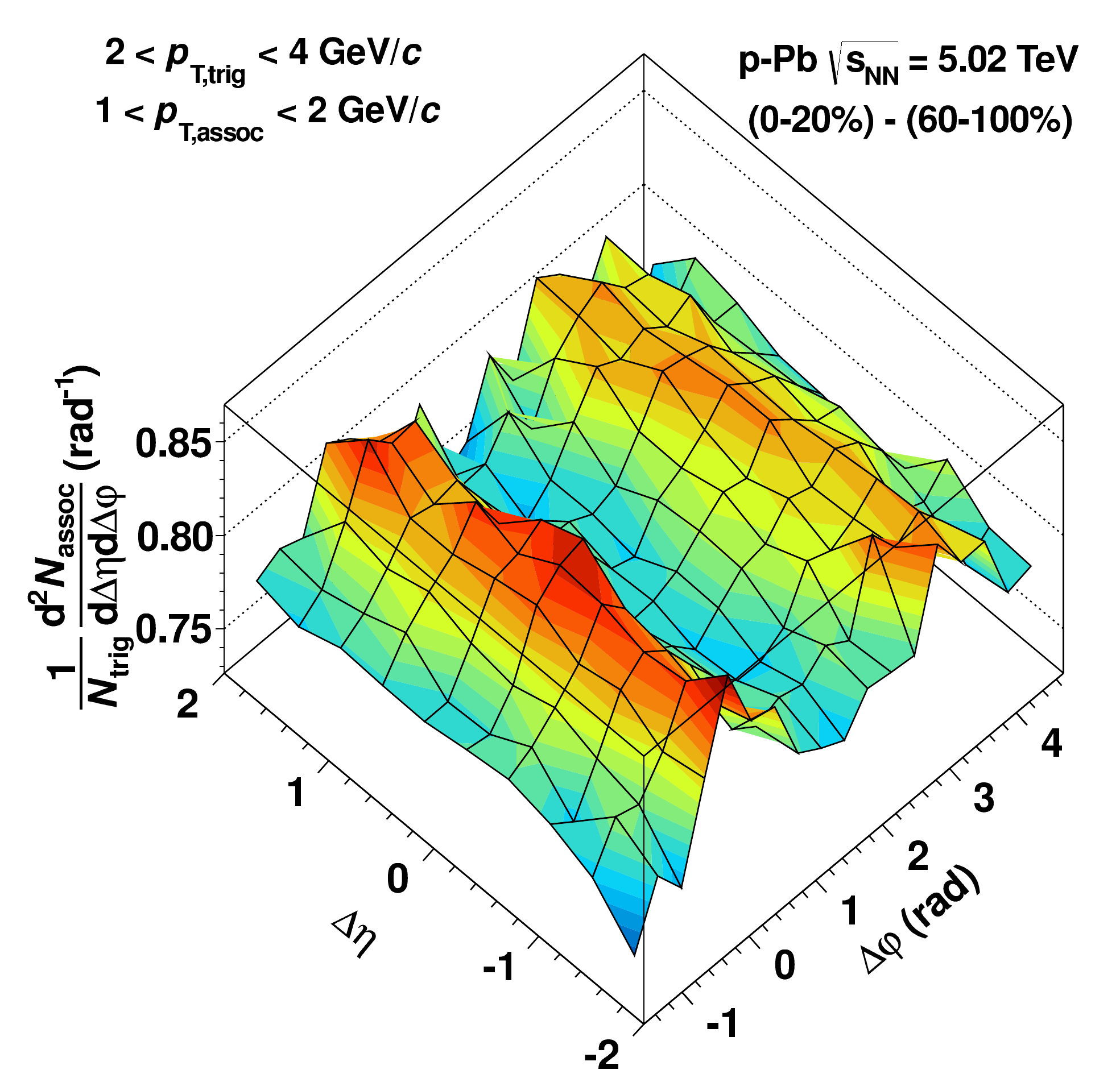}
 \caption{ The associated yield per trigger particle in $\Delta \phi$ and $|\Delta \eta|$ for pairs of charged particles 
 in p-Pb collisions. Left the associated yield for the 0--20\% multiplicity class and right after the 
 subtraction of the associated yield in the 60--100\% multiplicity class (figures taken from~\cite{Abelev:2012ola}).
  \label{fig:pPbcorr}. 
  }
\end{figure}
In Fig.~\ref{fig:pPbcorr} (left) the $\Delta \varphi$ $\Delta \eta$ correlations measured by ALICE in p-Pb collisions at 
$\sqrt{s_{_{NN}}} = 5.02$ TeV are plotted~\cite{Abelev:2012ola} 
for the events in the multiplicity class 0--20\%. 
A clear ridge structure at $\Delta \varphi = \pi$ (the away side ridge) 
is observed, as was also already seen in very high multiplicity p-p collisions by CMS~\cite{Khachatryan:2010gv}. 
At $\Delta \varphi = 0$ a clear jet-like peak is observed, which is, however, not isolated but situated on top of a ridge 
as well (the so-called near-side ridge). 
This near-side ridge is clearly visible in high multiplicity p-Pb collisions~\cite{Abelev:2012ola,CMS:2012qk, Chatrchyan:2013nka}. 
To quantify the change from low to high-multiplicity events, the correlated per trigger yield of the lowest multiplicity class is 
subtracted from the highest multiplicity class. The remaining distribution is shown in Fig.~\ref{fig:pPbcorr} (right), and exhibits 
a clear double ridge structure. These two ridges, one on the near-side and one on the away-side, 
are very similar~\cite{Abelev:2012ola,Aad:2012gla}.

Theoretically the origin of the observed ridge structure in p-Pb collisions is still the subject of 
speculation~\cite{Bozek:2012gr, Dusling:2012wy, Dumitru:2013tja, Basar:2013hea}. 
Currently, three theoretical proposals exist which might explain (part of) the observed azimuthal correlations. 
The first proposal is that the energy density achieved in p-Pb collisions 
is high enough that hydrodynamics using a Lattice QCD EoS can be used to describe the underlying physics. 
In that case, the spatial anisotropies in the initial state of a p-Pb collisions would generate anisotropic flow, 
resulting in significant values of $v_2$ and $v_3$~\cite{Bozek:2012gr}. 
The second proposal is inspired by the Color Glass Condensate 
description and claims that the ridge originates from collimated, in relative azimuthal angle, two-gluon 
production~\cite{Dusling:2012wy}. 
The third explanation is 
a combination of the previous two which invokes an CGC initial state with a finite number of sources which define 
the initial spatial anisotropy~\cite{Dumitru:2013tja}.

While these theoretical models can explain many of the observations and, in fact, even predicted these double ridge 
structures~\cite{Bozek:2012gr}, it is currently not clear if one can do an apples to apples comparison 
between theory calculations and experimental data. 
Most of the comparisons rely on the
fact that we either compare events with similar collision centrality 
(and some assumed relation with the initial spatial distribution) 
or compare events with a similar number of sources of particle production. 
In p-Pb, the correlation between the collision geometry or the number of sources to the measured multiplicity, which is used to
bin the experimental data, is under investigation and currently not well established. 
Therefore, in the remainder of the section, we 
focus only on the question if in experimental data the ridge observed in two-particle correlations is 
consistent with expectations of anisotropic flow. If this is the case, it should result in a well defined behaviour of the 
multi-particle cumulants and a clear mass hierarchy in the $\pt$-differential $v_2$ in p-Pb.  

\begin{figure}[htb]
 \includegraphics[width=0.3\textwidth]{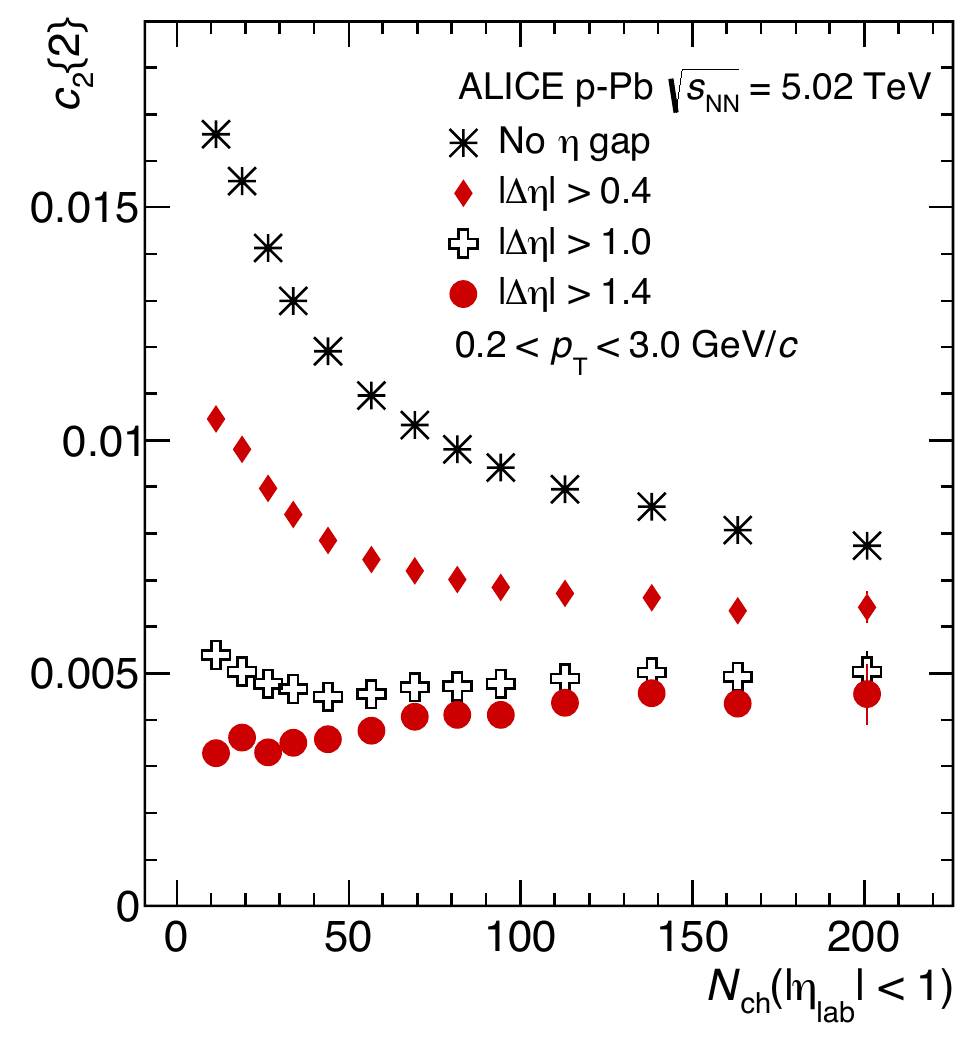}
 \includegraphics[width=0.34\textwidth]{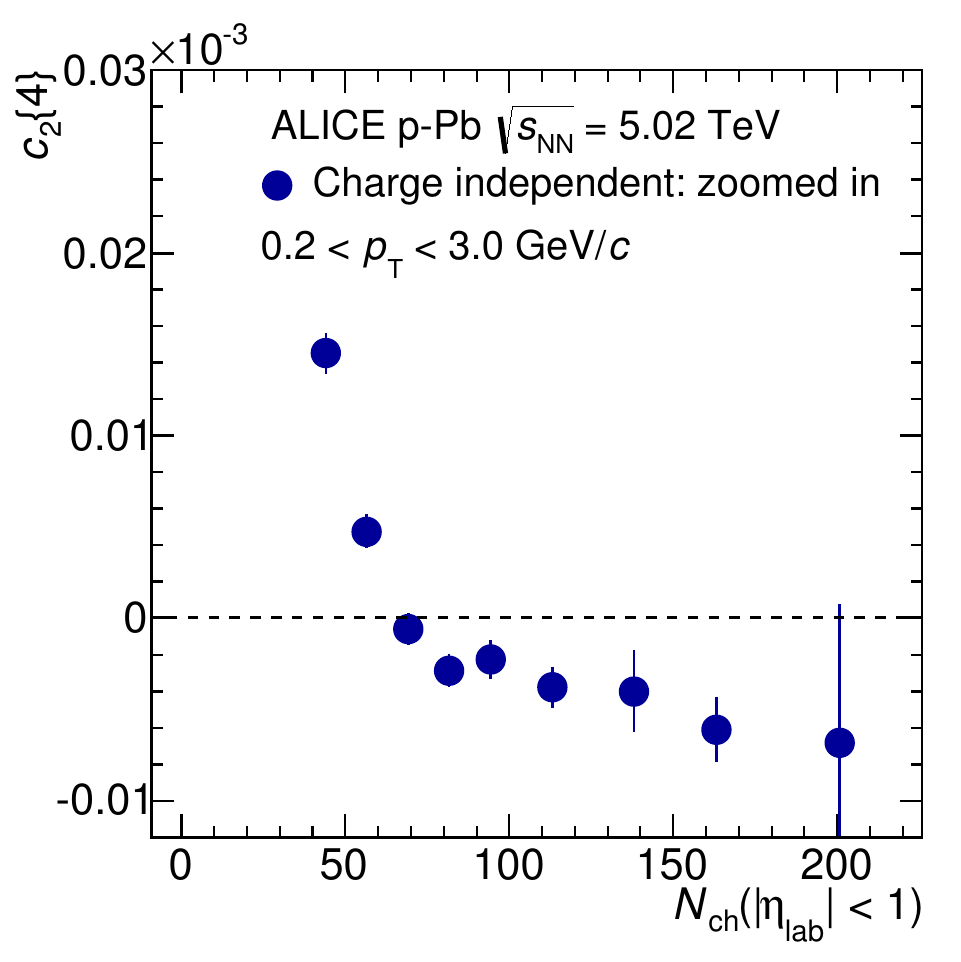}
 \includegraphics[width=0.34\textwidth]{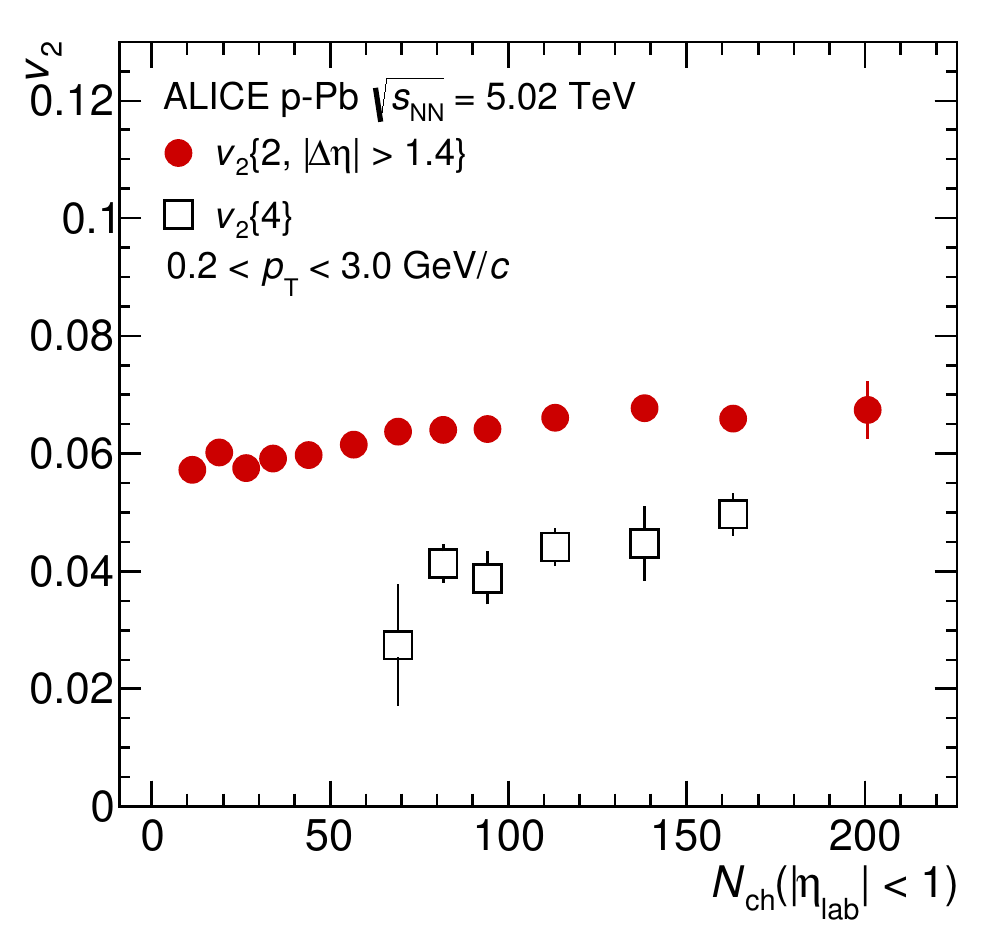}
 \caption{
 Two and multi-particle azimuthal correlations as function of the charged particle multiplicity in p-Pb collisions 
 measured by ALICE. 
      Left: The two-particle cumulant $c_2\{2\}$ for four different gaps in pseudo-rapidity.
      Middle: The four-particle cumulant $c_2\{4\}$.
      Right: The corresponding $v_2\{2\}$ and $v_2\{4\}$ (figures taken from~\cite{Abelev:2014mda}).
  \label{fig:pPbcumulants} 
  }
\end{figure}
The two-particle cumulant is plotted as a 
function of the charged particle multiplicity in p-Pb in the left panel 
of Fig.~\ref{fig:pPbcumulants} for four different gaps in pseudo-rapidity. 
When there is no gap in pseudo-rapidity between the particles the jet contribution clearly 
dominates and the two-particle cumulant decreases strongly with increasing multiplicity. This is expected due to simple 
combinatorics.  With increasing gap in pseudo-rapidity the multiplicity dependence clearly changes.
For the two-particle cumulant with $|\Delta \eta| > 1.4$  an increase with multiplicity is observed, which is compatible with 
the suppression of the jet-like peak in Fig.~\ref{fig:pPbcorr} (left) and the observed increase of the double ridge with 
increasing event multiplicity.

To investigate if this two-particle correlation results from a multi-particle correlation we plot 
in Fig.~\ref{fig:pPbcumulants} (middle) the four-particle cumulant as function of multiplicity. For the 
four-particle cumulant no gap in pseudo-rapidity is applied. For the events with a lower multiplicity the four-particle 
cumulant is positive which is incompatible with a dominant contribution coming from anisotropic flow. 
For events with a multiplicity of about 70 and higher the four-particle cumulant becomes negative, which is a 
prerequisite to determine the corresponding $v_2\{4\}$. 

In the right panel of Fig.~\ref{fig:pPbcumulants} the $v_2\{2\}$ and $v_2\{4\}$ are plotted versus multiplicity. 
The $v_2\{2\}$ is larger than $v_2\{4\}$ for the whole multiplicity range, similar to what is observed in 
Pb-Pb collisions~\cite{Abelev:2014mda,Aad:2013fja, Chatrchyan:2013nka}.
The difference between $v_2\{2\}$ and  $v_2\{4\}$ is indicative of anisotropic flow fluctuations but has also clearly a 
significant nonflow contribution. Measurements of $v_2\{6\}$ and $v_2\{8\}$ also have been performed in p-Pb 
collisions at the LHC by CMS~\cite{QM2014} (not shown here) and are found to be compatible with $v_2\{4\}$. 
In addition also a significant value of $v_3$ was observed.
Within current uncertainties all this is 
consistent with a elliptic power probability distribution~\cite{Yan:2014afa} of a limited number of sources in the initial spatial distribution 
$\varepsilon$~\cite{Yan:2013laa,Yan:2014afa}.

\begin{figure}[htb]
 \includegraphics[width=0.43\textwidth]{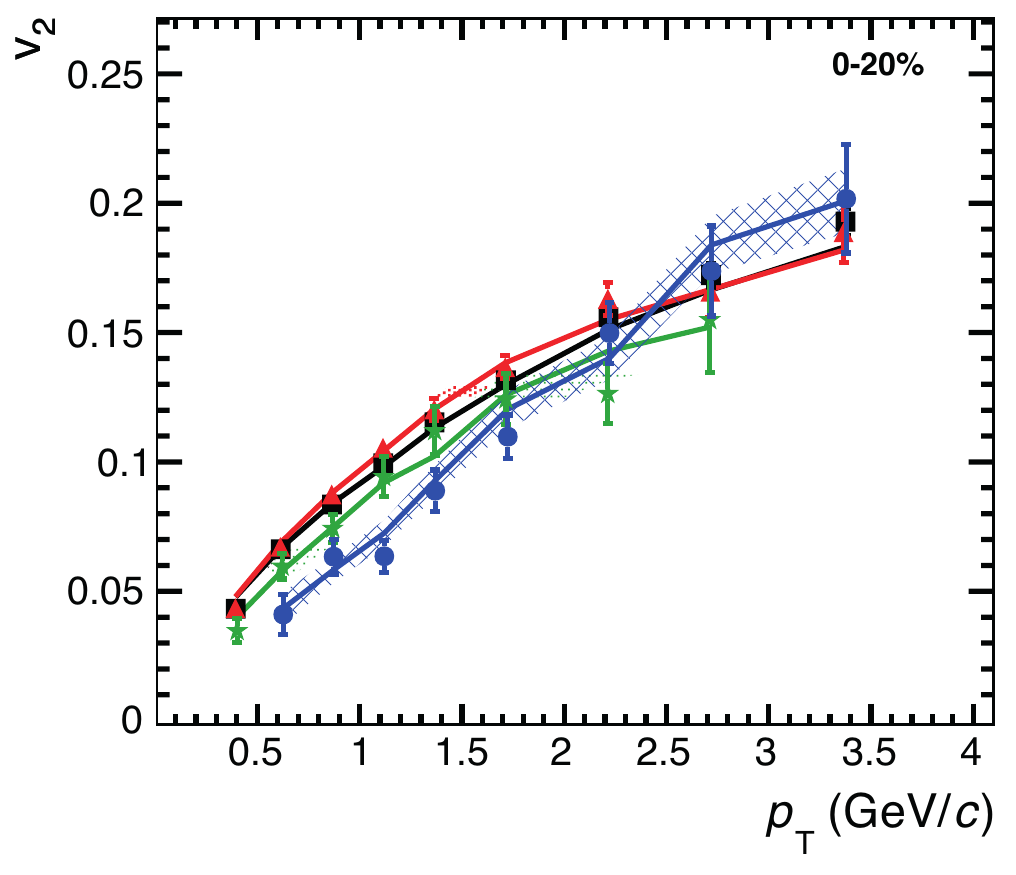}
 \includegraphics[width=0.57\textwidth]{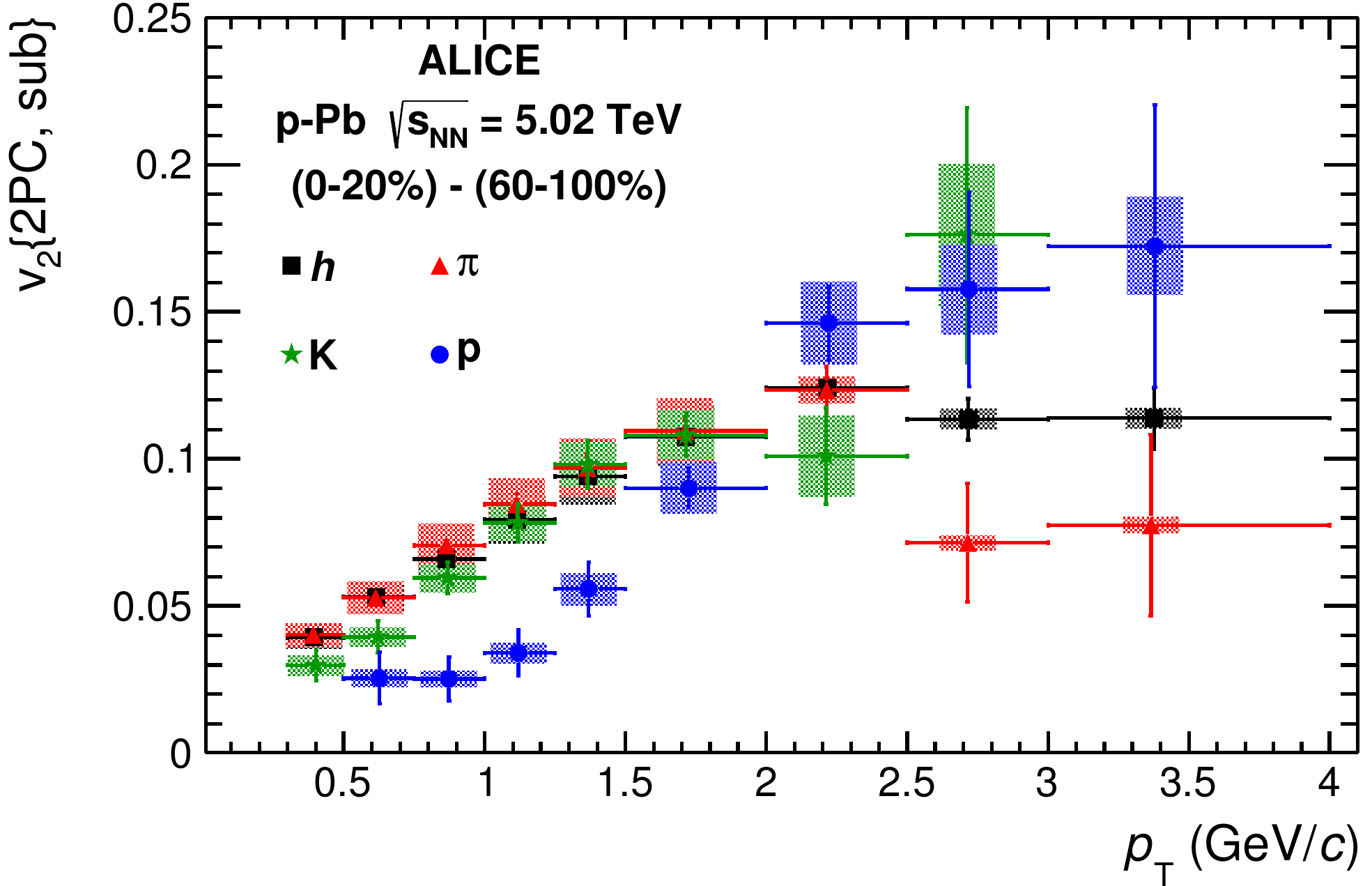}
 \caption{ Left: $\pt$-differential $v_2$ for all charged particles, pions, kaons and protons in p-Pb collisions for the 
 0--20\% multiplicity class. The solid lines are the $v_2$ obtained from particles correlated at fixed $\pt$ with particles from the full $\pt$ range. The symbols are the $v_2$ obtained from correlating particles in the same $\pt$ interval. 
Right: The $\pt$-differential $v_2$ for the 0--20\% multiplicity class after subtracting the per-trigger yield of 
the 60--100\% multiplicity class.
(figures taken from~\cite{ABELEV:2013wsa}). 
  \label{fig:pidpPb} 
  }
\end{figure}
The $\pt$-differential $v_2$ for different particles exhibits a characteristic mass hierarchy in Pb-Pb collisions, 
which in hydrodynamics is understood due to a mass dependent blueshift which is stronger in the flow plane.
If the ridge structures observed in p-Pb collisions also originate from a hydrodynamic expansion we would expect 
to also observe a mass hierarchy in $v_2(\pt, m)$ for these collisions. 
In Fig.~\ref{fig:pidpPb} the $\pt$-differential $v_2$ is plotted 
for the 0--20\% multiplicity class. The left panel shows the results for the charged particles as well as for the 
pions, kaons and protons separately. The results were obtained using a pseudo-rapidity gap of $|\Delta \eta| = 0.8$.
The $v_2$ are calculated correlating the particles from a certain $\pt$ interval 
with particles from the full $\pt$ range (solid lines), 
and compared to another procedure were both particles are taken from the same $\pt$ interval (symbols).
These results should agree if the measurement is dominated by correlations of each of the particles 
with a common flow plane and the $v_n$ fluctuations are $\pt$ independent. We observe that for the 0--20\% 
multiplicity class the lines and symbols are in agreement and that a clear mass hierarchy is 
observed~\cite{ABELEV:2013wsa}.

To again quantify the change from low to high-multiplicity events, the right panel of Fig.~\ref{fig:pidpPb} shows the 
resulting $v_2(\pt, m)$ after subtracting the per-trigger yield of the 60--100\% multiplicity class from that in the 0--20\% 
multiplicity class. After this subtraction the mass hierarchy becomes even more pronounced.

The correlation measurements in p-Pb show clear evidence for multi-particle correlations and are in qualitative agreement 
with hydrodynamical expectations. However, currently it is far from clear what the underlying microscopic mechanism is 
which is responsible for this behaviour.

\section{SUMMARY AND OUTLOOK}
\label{summary}

Anisotropic flow studies at the LHC have already produced a wealth of new results and many of the measurements
in Pb-Pb collision at LHC energies are relatively well understood in terms of a strongly coupled quark-gluon liquid 
as modelled in state of the art hydrodynamic model calculations. 
The initial density profile and its event-by-event fluctuations, which are currently the largest 
uncertainties in these calculations, are in particular much better understood due to the detailed $v_n$ measurements. 
At the same time there are also still many open questions which need to be addressed to reach the goal of 
a quantitative reliable description of heavy-ion collisions: 
\begin{itemize}
\item
The topic which currently gets considerable theoretical attention is the contribution to the flow from the very 
early stage (initial velocity profiles)~\cite{Becattini:2007sr,Vredevoogd:2008id,Heller:2012km,Casalderrey-Solana:2013aba,
vanderSchee:2013pia,Fries:2014oya}.
\item
In addition to the contribution to the flow from the very early stage there are also still open questions about 
the contribution from the late hadronic stage~\cite{Luzum:2013yya} and due to the particle production
mechanism~\cite{Molnar:2003ff}.
\item
Another hotly debated topic is if some of the angular correlations observed in p-p and p-Pb collisions, 
which traditionally are used as a reference for nucleus-nucleus collisions, are also related to the initial geometry.
\item
A topic discussed in other contributions to this volume is the flow of rare probes. 
It is important to see how the heavy quarks participate in the 
flow~\cite{ALICE:2013xna,Abelev:2013lca,Abelev:2014ipa,Uphoff:2011ad,Fochler:2011en,Batsouli:2002qf,Greco:2003vf,Liu:2009gx} 
but perhaps even more interesting is understanding the large flow observed for direct 
photons~\cite{Adare:2011zr,Lohner:2012ct,Chatterjee:2005de,Shen:2013cca,Shen:2014cga}.
\end{itemize}

While most of the measured $v_n$ were discussed in this review, the first harmonic $v_1$ was notably missing. 
The $v_1$ is sensitive to different 
interesting physics contributions~\cite{ATLAS:2012at,Abelev:2013cva,Teaney:2010vd,Luzum:2010fb,Gardim:2011qn,Retinskaya:2012ky,Snellings:1999bt,Csernai:1999nf} that, unfortunately, are difficult to disentangle. 
In this volume one review article focusses mainly on this harmonic~\cite{Csernai:2014cwa}. 

Finally, because flow is closely related to the geometry of the collision, a by-product of the flow analysis methods is 
the access to important orientation axes in a collision, like the orientation of the reaction-plane spanned by the 
beam axis and the impact parameter that connects the centre of the colliding nuclei. 
Exploration of the sensitivity of observables to the reaction- or flow-planes opens the road to very interesting physics analyses
like the detailed investigation of jet energy loss~\cite{Abelev:2012di,Snellings:1999gq,Bass:2008rv} 
and the possible effects of the large magnetic fields generated 
by the colliding nuclei~\cite{Abelev:2012pa,Abelev:2009ac,Kharzeev:2007jp,Fukushima:2008xe,Voloshin:2004vk,Gursoy:2014aka}.

\bigskip

\noindent
{\bf Acknowledgements:} I would like to thank Ante Bilandzic, Alexandru Florin Dobrin and You Zhou for the useful comments. 
The work of RS was partially supported by the Stichting voor Fundamenteel Onderzoek der Materie (FOM),  Nederlandse Organisatie voor Wetenschappelijk Onderzoek (NWO), and the Gastprogramm des Bayerischen Wissenschaftministerium.

\bigskip




\end{document}